\documentclass[%
 reprint,
 amsmath,amssymb,
 aps,
]{revtex4-2}

\usepackage{graphicx}
\usepackage{dcolumn}
\usepackage{bm}

\usepackage[caption=false]{subfig}
\usepackage{lipsum}
\usepackage{xcolor}
\usepackage{adjustbox}
\usepackage{booktabs} 
\usepackage{multirow, units}
\usepackage{placeins}
\usepackage{textgreek}
\usepackage{lineno}

\usepackage{physics}
\usepackage{upgreek}
\usepackage[colorinlistoftodos]{todonotes}

\let\originalzeta\zeta
\renewcommand{\zeta}{{\mathchoice
  {\scalebox{0.8}{$\displaystyle\originalzeta$}} 
  {\scalebox{0.8}{$\textstyle\originalzeta$}}     
  {\scalebox{0.8}{$\scriptstyle\originalzeta$}}   
  {\scalebox{0.8}{$\scriptscriptstyle\originalzeta$}} 
}}

\newcommand{\NF}[1]{%
  \overset{\smash{\raisebox{-0.2ex}{\hspace{0.1em}\scalebox{0.4}{$\boldsymbol{\frown}$}}}}{#1}%
}

\usepackage{tikz}

\usepackage{mwe}
\newcommand{\subfloattitle}[1]
{\makebox[\tempwidth][c]{\textbf{#1}}}

\begin{document}
\title{Numerical evaluation of the integrals of motion\\ in particle accelerator tracking codes}

\author{P.~B\'elanger\textsuperscript{1,2}, G.~Sterbini\textsuperscript{1}}

\affiliation{\vspace{2mm}
\textsuperscript{1}CERN, Geneva, Switzerland \\
\textsuperscript{2}TRIUMF, Vancouver, Canada\\
}

\date{\today}

\begin{abstract} 
\noindent Particle tracking codes are one of the fundamental tools used in the design and the study of complex magnetic lattices in accelerator physics. For most practical applications, non-linear lenses are included and the Courant-Snyder formalism falls short of a complete description of the motion. Likewise, when the longitudinal motion is added, synchro-betatron coupling complicates the dynamics and different formalisms are typically needed to explain the motion. In this paper, a revised formalism is proposed based on the Fourier expansion of the trajectory --- known to be foundational in the KAM theorem --- which naturally describes non-linear motion in 2D, 4D and 6D. After extracting the fundamental frequencies and the Fourier coefficients from tracking data, it is shown that an approximate energy manifold (an invariant torus) can be constructed from the single-particle motion. This cornerstone allows to visualize and compute the areas of the torus projections in all conjugate planes, conserved under symplectic transformations. These are the integrals of motion, ultimately expressed in terms of the Fourier coefficients. As a numerical demonstration of this formalism, the case of the 6-dimensional Large Hadron Collider (LHC) is studied. Examples from the 2D and 4D Hénon map are also provided. Even for heavily smeared and intricate non-linear motion, it is shown that invariant tori accurately describe the motion of single particles for a large region of the phase space, as suggested by the KAM theorem.
\vspace{5mm}
\end{abstract}

\maketitle


\section{Introduction}
Particle accelerators and storage rings such as the Large Hadron Collider (LHC) at CERN require their beam of particles to be stable for long periods of time. In the LHC, the circulating beams can be stored for dozens of hours (57~h achieved in a single fill in 2022). This corresponds to more than $10^9$ revolutions around the accelerator --- a number comparable to that of Earth's revolutions around the Sun over the past few billion years. As such, the $10^{14}$ protons of the LHC beam experience a comparable journey to that of our Solar System, \textit{daily}. This machine therefore offers an ideal testing ground to study the stability of conservative Hamiltonian systems, historically tied to the famous \textit{$n$-body problem} in celestial mechanics. In its simplest formulation, the Kepler problem, two masses orbit each other and the motion remains stable for all eternity. Adding additional masses, say by considering Jupiter alongside the Earth and the Sun, brings us into the famous three-body problem, which has no general closed-form solution. The system can then exhibits chaotic motion and the question of stability becomes a complicated one. This is what led J.~Moser to ask~\cite{moser_is_nodate}: \textit{Is the solar system stable?} And similarly, one could reasonably ask: \textit{Is the LHC beam stable?}\\

That being said, most practical applications do not require us to know if the motion is stable for all eternity, but simply for a limited \textit{meaningful} period of time, which lifts a lot of mathematical subtleties and opens the door for numerical methods to be used. When the single-particle motion is not stable, or \textit{regular}, it is said to be \textit{chaotic}, in the sense of Poincaré~\cite{alma9913588812806531,moser_is_nodate}, which is to say that the trajectory cannot be represented as a convergent series. Chaotic motion  may appear to be random at all times, or it may appear to be regular over long periods of time before diverging~\cite{percival_chaos_1987}. If this characteristic time length is longer than the lifetime of the circulating beams, then such a trajectory might be, by all practical accounts, indistinguishable from a stable one.\\

In any case, the behaviour of a system as a whole is in general neither completely regular nor chaotic, but \textit{mixed}. As demonstrated by the KAM theorem~\cite{moser_is_nodate} (1954-1963), mixed systems always admit some regular solutions close to integrable regions of the system. This can be seen, for example, with the gaps in the rings of Saturn or the gaps in the observed frequencies of the asteroids orbiting our Sun~\cite{laskar_large_nodate}: unstable chaotic trajectories amidst stable ones. With the popularization of computational physics, both types of trajectories can be observed and studied, although the challenge presented by chaos still remains. To paraphrase S.~Wolfram~\cite{wolfram2023,zwirn_unpredictability_2013}: ``We've come to believe that there would be formulas to predict everything. But computational irreducibility [\textit{e.g.}, for chaotic systems] shows us that that isn't true, and that in fact, to find out what a system will do, we have to go through the same irreducible computational steps as the system itself". As such, chaotic motion is perfectly deterministic, but the future state of the system cannot be found readily from its present state. This simple fact establishes \textit{particle tracking} as the fundamental tool to be used in order to better our understanding of complex machines like the LHC.

\subsection*{Particle Accelerators}
By design, the magnetic lattice of storage rings is built mainly out of linear elements: dipolar and quadrupolar magnets for which the Hamiltonian admits closed-form solutions (analogous to the Kepler problem). Of course, real machines make use of higher order magnets (sextupoles, octupoles, etc.) and feature high order multipolar contributions such that the final non-linear system admits mixed solutions. Despite this complexity, the linear picture has become the standard description of accelerator physics since the work of Courant and Snyder in 1958~\cite{courant_theory_1958}. In fact, the Courant-Snyder theory~\cite{wolski_beam_2014} has been a cornerstone in the design, description, and operation of particle accelerators. The success of this approach is partly due to the fact that its main assumptions --- that of linear and uncoupled dynamics --- have proven to be excellent approximations to describe realistic stable machines.\\

However, as highlighted by A.~Chao~\cite{chao_slim_nodate}, the Courant-Snyder picture is not without weakness and fails to describe non-linear dynamics in a natural way. Indirectly, the ubiquity of this linear formalism popularized the idea that elliptical phase space topologies are to be expected in particle accelerators. With non-linear magnets present in the lattice, this formalism is of course unsatisfactory and one should instead expect the particle trajectories to follow a more complex shape, described by a more general quasiperiodic expansion (of which ellipses are only a particular case~\cite{belanger_topo_2024}).
In fact, this was already well described by the epicycle theory of the Greeks, formalized in the \textit{Almagest of Ptolemy}~\cite{gallavotti_quasi_1999}, which is to say that the motion can be decomposed into a finite number of uniform circular motions. This age-old claim still finds itself to be useful in modern times, as discussed in later sections.\\

On the analytic front, these ideas are not so far from the picture provided by the \textit{normal form} approach~\cite{bazzani_normal_1994-2}, which aims at finding an approximate interpolating Hamiltonian to describe the system. In the last few decades, this framework allowed for the formulation of the so-called resonance driving terms (RDTs), which are used to describe perturbations from the linear behaviour, arising from the presence of high order multipolar terms and other spurious effects~\cite{franchi_first_2014}. Correction schemes based on experimental measurements of resonance driving terms have had important practical results for the successful operation of the LHC. That being said, when the machine complexity is increased, far-reaching assumptions are needed for those methods to succeed, without which the analytic calculations rapidly become intractable.\\

\newpage
One can therefore understand the interest that was put, over the years, towards the development of \textit{symplectic} particle tracking codes~\cite{forest_geometric_2006-1}, driven by the need to accurately investigate the complex dynamics of non-linear systems beyond the reach of analytic calculations. By numerically implementing symplectic elements~\cite{dragt_overview_2013-1}, tracking codes can be used to build virtual machines which closely resemble the real machines being modelled and for which the single-particle physics is accurate. From there, complex phenomena such as particle diffusion or chaos can be studied via brute-force tracking of many particles over long periods of time. But as mentioned, when trying to interpret the results, the linear formalism commonly used in accelerator physics falls short of a proper description of the non-linear effects. \\

The present paper is set to address the aforementioned limitations. It needs to be said that no new mathematical results are presented, most of which can be found in earlier work~\cite{bazzani_analysis_1997}. Instead, we propose a revised formalism which allows for a deeper conceptualization and understanding of non-linear motion in particle accelerators, both for the longitudinal and the transverse motion. First, it is shown (section~\ref{sec:theory}) that the quasiperiodic expansion describing the time series of regular trajectories can be obtained numerically --- from tracking --- to a high level of accuracy. From there, the approximate energy manifold of the particles can be constructed following a simple change of variable. This topological jump (section~\ref{sec:energy_manifold}), in turn, allows for the visualization and the computation of an   area which is conserved by the symplectic transformations of the system --- the so-called integrals of motion --- in 2D, 4D and 6D (section~\ref{sec:integrals_of_motion}). Additionally, coupled motion and phase space smearing appear naturally within the context of this formalism. Finally (section~\ref{sec:application_LHC}), these concepts are applied to tracking data from the 6D LHC, where it is shown that the integrals of motion can indeed be accurately computed for a large region of the phase space.


\section{Theory \& Formalism \label{sec:theory}}
The Hamiltonian flow of a conservative system (\textit{e.g.} a non-dissipative storage ring) can be expressed via a symplectic \textit{map}, which describes the discrete evolution from an initial set of coordinates to a final set of coordinates, some finite time later~\cite{meiss_symplectic_1992}. This is the essence of particle tracking codes, where single particles are transported along the accelerator and their trajectories are observed in phase space, turn after turn. It can be shown that a simple non-linear lattice made out of dipoles, quadrupoles and a single sextupole follows the well-known Hénon map~\cite{Henon_numerical_1969} (see Appendix~\ref{sec:Henon_map} for details of the map), which will be used for illustrative purposes in the following sections.

\newpage
\subsection{KAM Theorem}
A fundamental aspect to consider for the stability of single-particle trajectory is the question of integrability.  A system with $n$ degrees of freedom (a $2n$ dimensional phase space) is said to be \textit{integrable} if there exist $n$ independent integrals of motion $I_i$, also called \textit{actions}, which commute under the Poisson bracket $\{I_i,I_j\} = 0$ ($i,j = 1,2,\dots,n$). For a realistic particle accelerator with 3 degrees of freedom, one should therefore expect to be able to find 3 integrals of motion~\cite{Henon_applicability_1964} such that the Hamiltonian $\mathcal{H}$ can be written as a function of the actions only: 
\begin{equation}
    \mathcal{H}(x,p_x,y,p_y,\zeta,p_\zeta) \mapsto \mathcal{H}(I_x,I_y,I_\zeta)
\end{equation}

\begin{figure}[t!]
    \centering
    \includegraphics[width=0.8\linewidth]{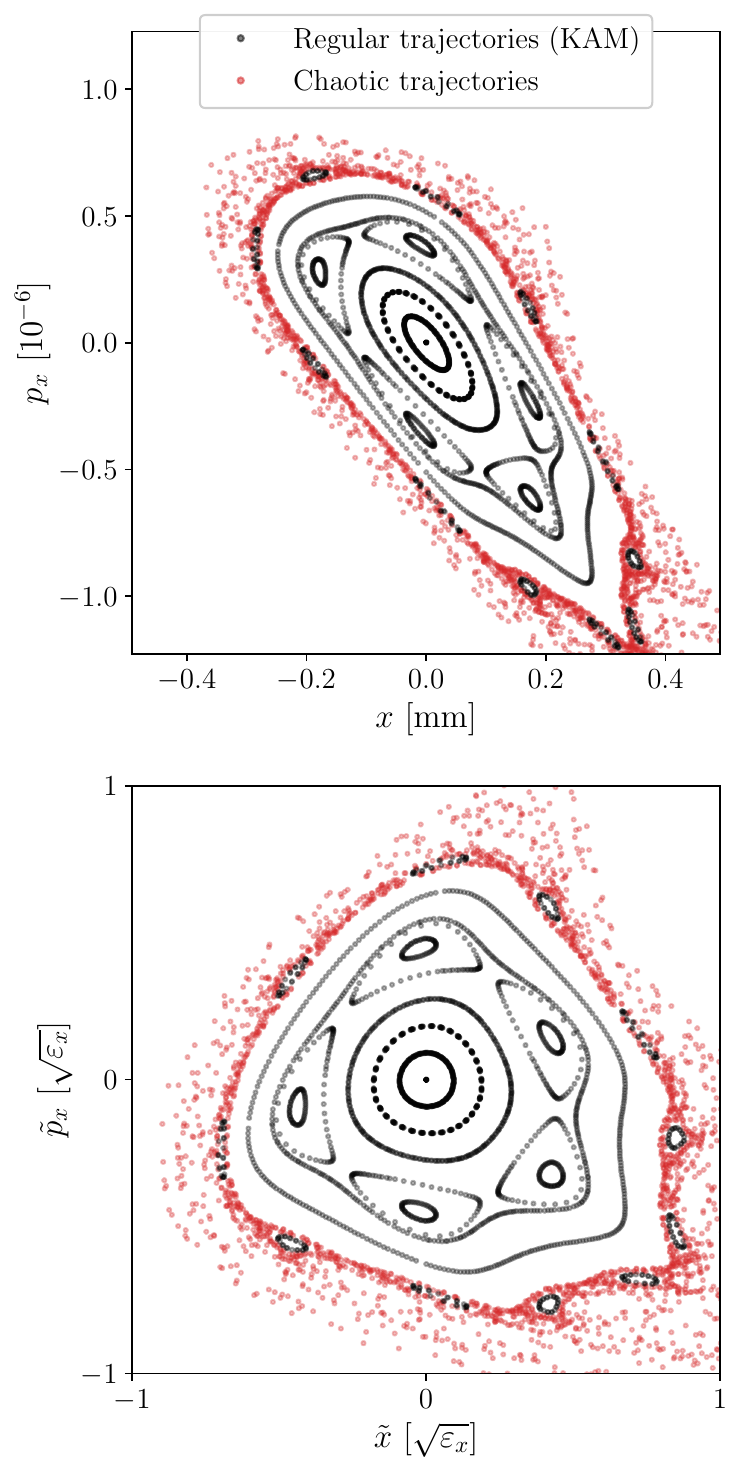}
    \caption{Phase space portrait of the 2D Hénon map ($\mu_x = 0.2071$) in the physical phase space with $\varepsilon_x = 0.335\ \text{nm}$ (top plot) and the Courant-Snyder phase space (bottom plot). Both regular and chaotic trajectories can easily be identified. The KAM theorem guarantees the existence of some regular trajectories near the elliptical fixed points.}
    \label{fig:Henon_KAM}
\end{figure}

\noindent In general, however, the Hamiltonian might be not integrable for all initial conditions and the system is mixed. It exhibits both regular motion, for which the series development of the trajectory converges, and chaotic motion, for which it does not. Fig.~\ref{fig:Henon_KAM} shows the phase space portrait of the Hénon map, for which the two types of trajectories can be clearly identified. As mentioned by M. Hénon, regular trajectories ``seem to lie exactly on a curve"~\cite{Henon_applicability_1964}, which takes the form of a quasiperiodic series (see following section). The KAM theorem, from Kolmogorov, Arnold and Moser, guarantees the \textit{existence} of some of those regular trajectories for certain classes of mechanical systems~\cite{moser_is_nodate,percival_chaos_1987}. This is notably the case of the $n$-body problem, as well as restricted regions of non-integrable systems where the Hamiltonian is almost integrable (in a perturbative sense). Particle accelerators belong to this latter category. Since particle accelerators are built by adding high-order multipoles to a linear foundation of dipoles and quadrupoles, they are intrinsically designed in a perturbative fashion. Hence, the final Hamiltonian is well integrable near the closed orbit --- stable, by construction --- of the machine and stable KAM solutions are guaranteed to \textit{exist}, up to some amplitude.

\subsection{Choice of Coordinates \label{sec:coordinates}}
Naturally, the single-particle motion at a given $s$ position along the ring can be described in the physical phase space $\vec x = (x , p_x, y, p_y,\zeta,p_\zeta)$ directly from tracking (see Appendix~\ref{sec:xsuite_coordinates} for details), as shown in the top plot of Fig.~\ref{fig:Henon_KAM}. However, to highlight the effect of non-linearities, the coordinates can always be transformed into the Courant-Snyder (or \textit{linearly normalized}) phase space coordinates $
\vec{\tilde x} =  (\tilde x , \tilde p_x, \tilde y, \tilde p_y,\tilde \zeta,\tilde p_\zeta)$ with the help of the $W$-matrix~\cite{cern_xsuite_nodate} using $\vec{\tilde x} = W^{-1}(s)\ \vec{ x}$. By doing so, linear motion (\textit{e.g.} close to the origin) takes the form of circles instead of ellipses, as shown in the bottom plot of Fig.~\ref{fig:Henon_KAM}. A convenient way to describe the turn-by-turn motion of a particle is to use the complex notation $\tilde \psi_x = \tilde x -i\tilde p_x$, combining the position and the momentum together~\cite{bartolini_normal_1997}. One can then write, for 2D linear motion:
\begin{align}
    \tilde \psi_x (N)=\tilde x - i\tilde p_x = A_0\  e^{i[2\pi Q_xN] }\quad  (\text{for}\ I_x\to 0)
    \label{eq:linear_motion}
\end{align}

\noindent where $N$ is the turn number, $|A_0| = \sqrt{2I_x}$ is the radius of the circle in the complex plane and $Q_x$ is the \textit{fundamental frequency} of the particle. In the Courant-Snyder phase space, the coordinates ($\tilde x$, $\tilde p_x$, etc.) homogeneously assume units of $\sqrt{\text{m}}$. If the beam emittance is known, one can further express the coordinates in units of $\sqrt{\varepsilon}$ such that $\tilde x = 1\ \sqrt{\varepsilon_x}\ \mapsto\ x = 1\ \sqrt{\beta_x \varepsilon_x} =1\ \sigma_x $, which relates to the beam size.\\

Regular trajectories can in principle be further simplified by going to \textit{normal form}, $
\vec{\NF x} =  (\NF x , \NF p_x, \NF y, \NF p_y,\NF \zeta,\NF p_\zeta)$, a phase space where the motion takes the form of purely independent rotations around a given fixed point (typically the origin) expressed in terms of the action-angle variables~\cite{bazzani_normal_1994-2} --- similar to eq.(\ref{eq:linear_motion}), but for all amplitudes. That being said, this transformation is not straightforward and cannot be done, in general, without resorting to numerical methods like the ones presented in this paper. It is nonetheless important to note that there \textit{exists}, in principle, such a transformation which can be used to normalize the motion step-by-step, going from the physical to Courant-Snyder to normal form phase space, $(x,p_x) \mapsto(\tilde x,\tilde p_x)\mapsto(\NF x,\NF p_x)$, in a symplectic manner.

\subsection{Regular Motion \& Quasiperiodic Expansion \label{sec:regular_motion}}
One important practical implication of the KAM theorem is to provide us with a closed-form expression to describe regular trajectories. Using the complex notation $\tilde \psi_x = \tilde x - i\tilde p_x$ to describe both position and momentum in the Courant-Snyder phase space, the motion in each plane can be written as three Fourier series, each made of $N_h$ harmonics, following:
\begin{equation}
\begin{aligned}
    \tilde \psi^{(N_h)}_{x}(N) &= \tilde x- i  \tilde p_x = \sum_{k=0}^{N_h} {A_k\ e^{i[2\pi(\vec n_k \cdot \vec Q)N] }}\\
    \tilde \psi^{(N_h)}_{y}(N) &= \tilde y- i  \tilde p_y = \sum_{k=0}^{N_h} {B_k\ e^{i[2\pi(\vec m_k \cdot \vec Q)N] }}\\
    \tilde \psi^{(N_h)}_{\zeta}(N) &= \tilde \zeta- i  \tilde p_\zeta = \sum_{k=0}^{N_h} {C_k\ e^{i[2\pi(\vec \ell_k \cdot \vec Q)N] }}\\
\end{aligned}
\label{eq:psi_x}
\end{equation}

\noindent giving us the coordinates as a time series in $N$, the turn number. The spectral amplitudes $A_k, B_k$ and $C_k$ are complex numbers, sorted with decreasing amplitudes for increasing $k$. In general, this description of the motion is only an \textit{exact} solution to the KAM problem when $N_h \to \infty$. By truncating the quasiperiodic
expansion to a finite $N_h$, one obtains an approximate solution where the error is periodic and bounded, provided that the KAM solution exists. It is noteworthy that the series development in each plane only contains contributions from a \textit{discrete set} of well-defined spectral lines $\nu_k$ (or frequencies, with units of turn$^{-1}$)~\cite{simo_introduction_1999}. For example, in the horizontal plane, $\tilde \psi_x$, one can write:
\begin{equation}
    \nu_k = \vec n_k \cdot \vec Q + n_{0k},\qquad \vec n_k \in \mathbb{Z}^3,\ n_{0k} \in \mathbb{Z}
    \label{eq:nu_k}
\end{equation}

\noindent where the $\vec n_k \cdot \vec Q = (n_x,n_y,n_\zeta)_k\cdot (Q_x,Q_y,Q_\zeta)$ are integer linear combinations (and so are $\vec m_k \cdot \vec Q$ and $\vec \ell_k \cdot \vec Q$) of the particle's fundamental frequencies, $\vec Q$, properties of the single-particle trajectory, shared between the three planes $\tilde \psi_x$, $\tilde \psi_y$ and $\tilde \psi_\zeta$. For the convenience of the graphical representation, the frequencies are aliased in the domain $\nu_k \in [-0.5,0.5]$ using the additional integer $ n_{0k}$, without loss of generality.\\

\begin{figure}[t!]
    \centering
    \includegraphics[width=0.97\linewidth]{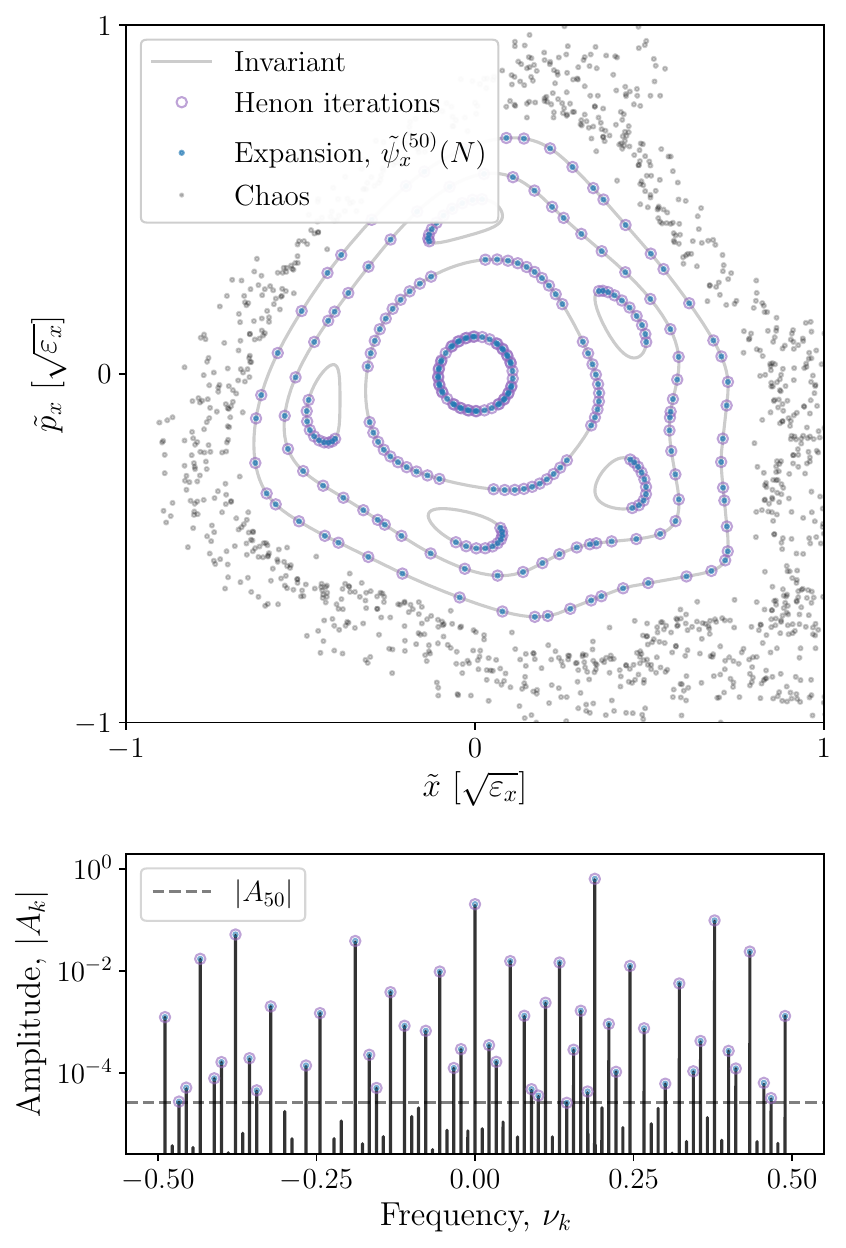}
    \caption{Comparison of the time series obtained from the quasiperiodic expansion of eq.~(\ref{eq:psi_x}) for the 2D Hénon map ($\mu_x = 0.2071$) and the map itself, shown for the last 50 turns out of $10^5$ iterations. The expansion is numerically obtained from $N_h =50$ harmonics evaluated after $10^4$ turns. The validity of the time series therefore extends outside of its evaluation window. The bottom plot shows the Fourier spectrum of the outermost stable particle from the plot above: only the discrete set of frequencies from eq.~(\ref{eq:nu_k}) are found.}
    \label{fig:Henon_phasor_expansion}
\end{figure}

In Fig.~\ref{fig:Henon_phasor_expansion}, the turn-by-turn position obtained from eq.~(\ref{eq:psi_x}) (truncated to $N_h = 50$ harmonics) is compared to the iterations of the Hénon map itself from which it was numerically extracted. As shown in the bottom plot, the Fourier spectrum of a regular trajectory contains a discrete set of frequencies, made out of linear combinations of the fundamental frequency --- one only, for this 2D case --- as stated by eq.~(\ref{eq:nu_k}). The description of the motion given by eq.~(\ref{eq:psi_x}) is an important cornerstone of the present approach, allowing for a closed-form analysis of the trajectories. One can see that the turn-by-turn position lies on an \textit{invariant} curve, which will be introduced in the next section. The area enclosed by this loop is an integral of motion, preserved under Hamiltonian transformations.\\
%


The interest of eq.~(\ref{eq:psi_x}) lies in the fact that the Fourier expansion can be obtained numerically from a Fourier analysis or  --- to a better accuracy --- from the Numerical Analysis of the Fundamental Frequencies (NAFF), introduced by Laskar~\cite{simo_introduction_1999}. This claim holds even for complex non-linear systems, where analytic calculations become impractical. The NAFF algorithm (see \texttt{nafflib}~\cite{nafflib_pypi}, or Appendix~\ref{sec:NAFF}~\cite{laskar_frequency_1993,laskar_measure_1992,laskar_frequency_2003,papaphilippou_detecting_2014}) is an iterative scheme where a windowing function is used to recover the frequencies of the signal with an accuracy that is several orders of magnitude better than a simple Fast Fourier Transform (FFT). \\

After finding the dominant frequency, $\nu_0$, its contribution is subtracted from the original signal and the procedure is repeated to obtain the second dominant frequency, $\nu_1$, and so on. By doing so, one eventually obtains the set of sorted amplitudes, $A_k$, and frequencies, $\nu_k$, corresponding to the quasiperiodic expansion of eq.~(\ref{eq:psi_x}) with great accuracy for a relatively low number of turns. For the 2D Hénon map shown in Fig.~\ref{fig:Henon_phasor_expansion}, $10^4$ turns are sufficient to get most frequencies up to machine precision, which allows to describe the time series of the quasiperiodic trajectory beyond the analysis window, say up to $10^5$ turns in this particular example. Evidently, the numerical description obtained is an approximation of the real trajectory of the map.

\section{Energy Manifold \label{sec:energy_manifold}}
For the non-dissipative conservative case, individual particles conserve their energy and the single-particle trajectories are constrained to evolve on well-defined constant-energy surfaces, called --- in phase space --- \textit{energy manifolds}. By construction, these energy manifolds are invariant under the one-turn map of the system, \textit{i.e.} the mapping of any point from the energy manifold also
belongs to the manifold itself, one turn later. This can be seen, for
example in 2D, from the invariant curves of the Hénon map
depicted in Fig.~\ref{fig:Henon_phasor_expansion}. The following sections generalize this idea to 4D and 6D systems and present a description which allows for the calculation of the integrals of motion.

\begin{figure*}[ht!]
    
    \centering
    \hfill
    \subfloat{\centering \large Courant-Snyder phase space}
    ~\hspace{0.24\textwidth}
    \subfloat{\centering \large Normal Form phase space}
    ~\hspace{0.05\textwidth}
    \hfill
    \medskip
    
    \subfloat{%
    \centering
    \includegraphics[width=0.49\textwidth]{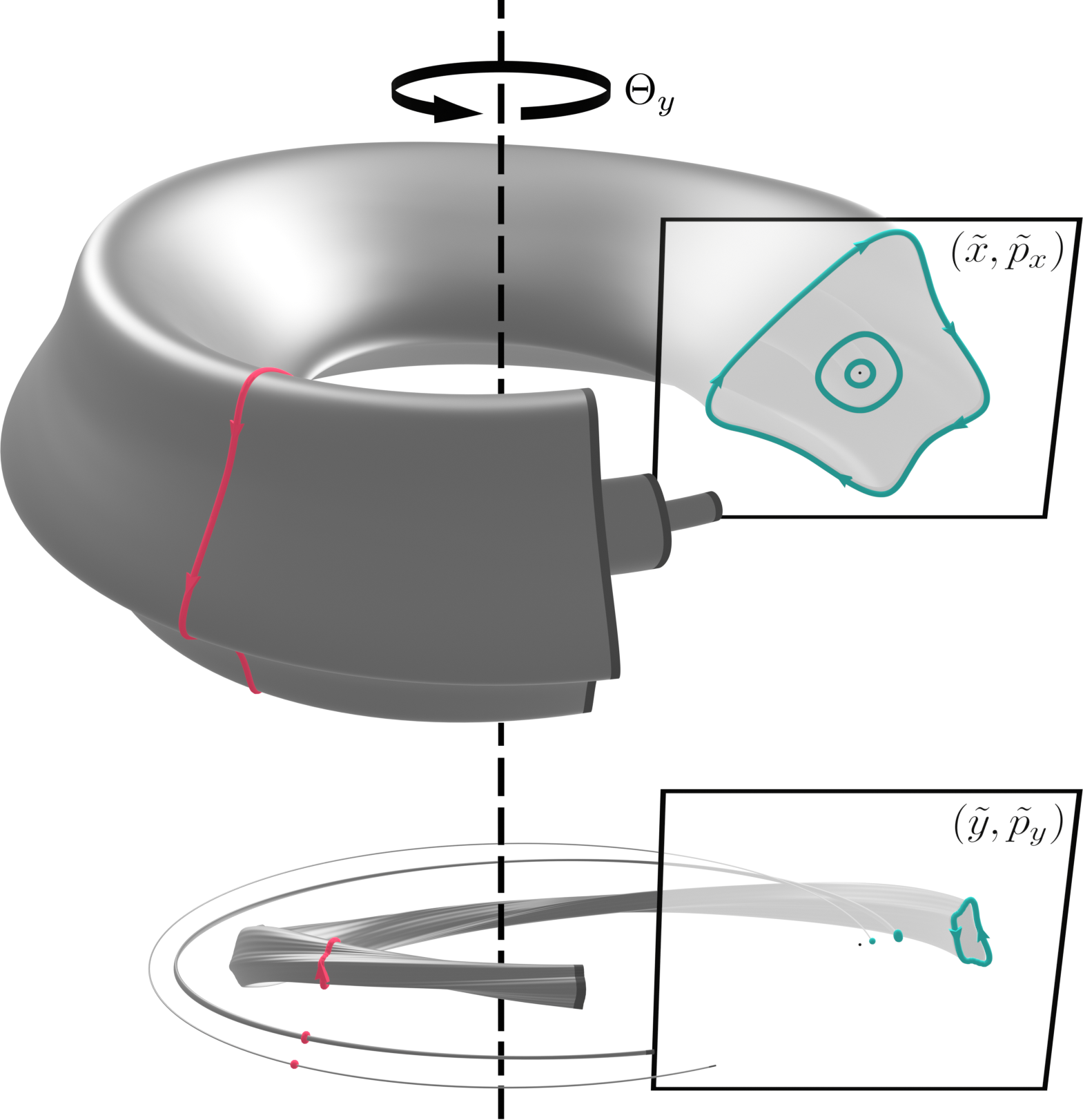}
    }
    ~\hspace{0.015\textwidth}
    \centering
    \subfloat{%
    \includegraphics[width=0.49\textwidth]{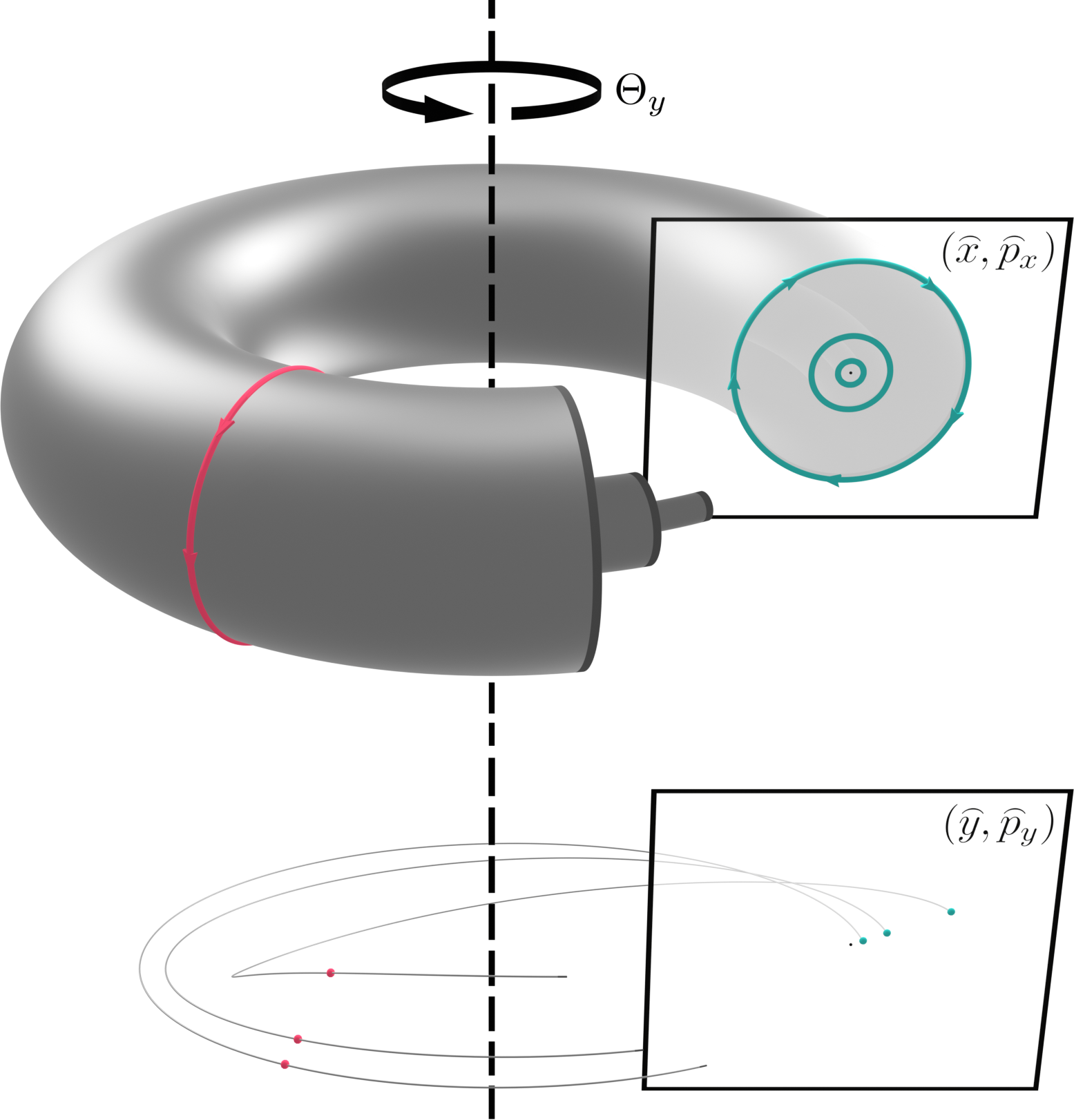}
    }
    \medskip 

    \subfloat{%
    \centering
    \includegraphics[width=0.49\textwidth]{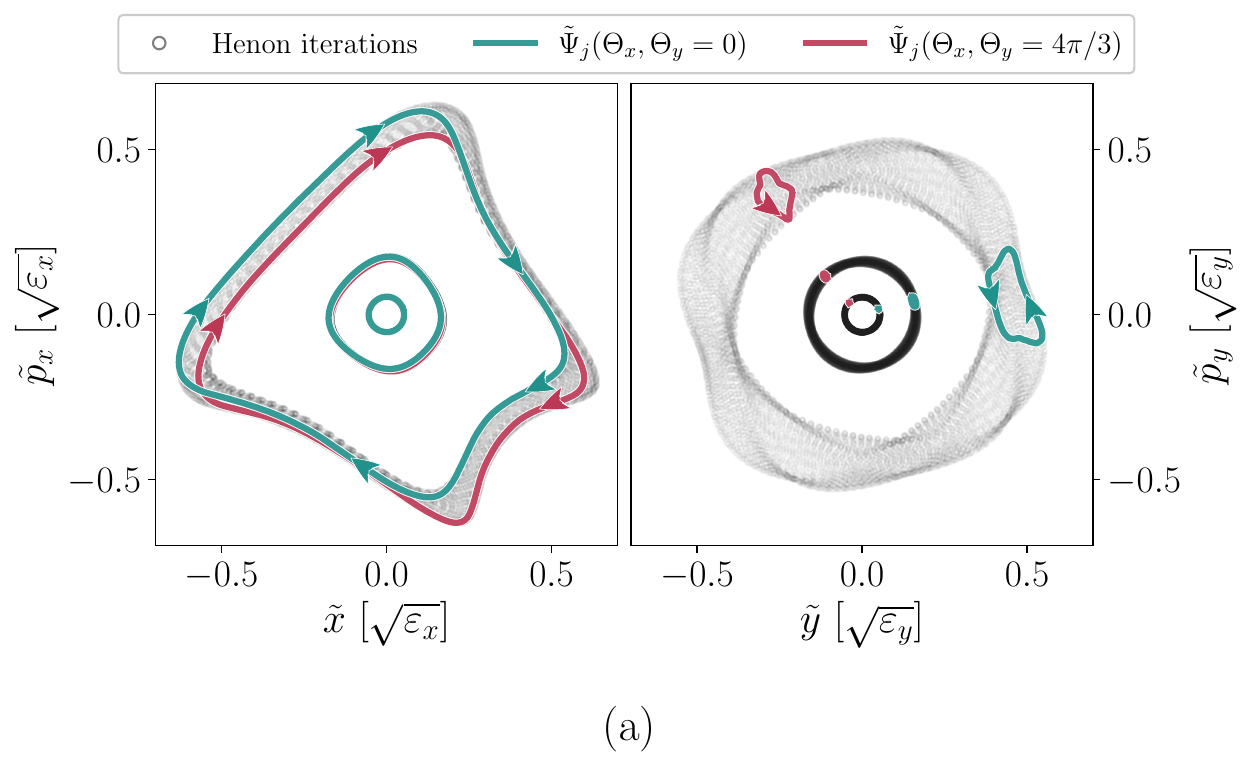}
    }
    ~\hspace{0.015\textwidth}
    \centering
    \subfloat{%
    \includegraphics[width=0.49\textwidth]{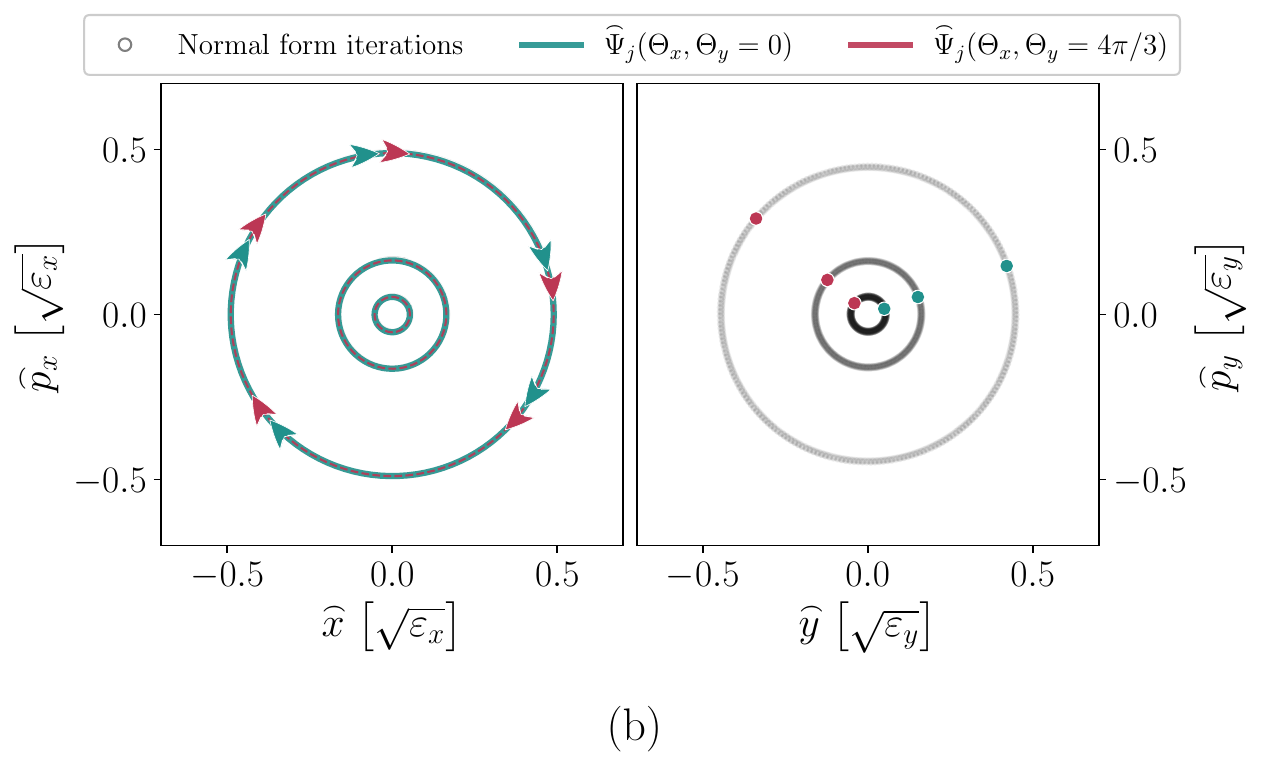}
    }
    
    \caption{Visualization of the energy manifolds $\tilde \Psi$ (invariant tori) from eq.~(\ref{eq:Psi_topology}), in grey, for 3 different particles in the 4D Hénon map with $\mu_x = 0.2465$, $\mu_y = 0.4142$ and $\rho = 0.3$ (see Appendix~\ref{sec:Henon_map} for details). (a) In the Courant-Snyder phase space, $\vec{\tilde x}$. (b) In the normal form phase space, $\vec{\NF x}$. The top plots show the projections of the invariant tori, $\tilde \Psi_x(\Theta_x,\Theta_y = \text{cst.})$ and $\tilde \Psi_y(\Theta_x,\Theta_y = \text{cst.})$, for all $\Theta_y$. The projections of two different 4-dimensional $\Theta_x$ loops are also shown in red and green, travelled following the directional arrows. The area is positively oriented if travelled clockwise. The sum of the areas enclosed by corresponding colored loops is an invariant. The azimuthal deformation of the tori gives rise to the smearing observed in the single-particle trajectory over many turns, when plotted in the conjugate planes, as shown in the bottom plots.}
    \label{fig:4D_torus}
\end{figure*}

\subsection{Invariant Tori}
By following a given particle (\textit{e.g.} in Fig.~\ref{fig:Henon_phasor_expansion}) over many turns using eq.~(\ref{eq:psi_x}), one finds that it eventually covers the invariant curve --- the energy manifold --- on which it resides \textit{densely}. Out of the quasiperiodic turn-by-turn motion emerges an underlying periodic structure. This is to say that the main argument of the quasiperiodic expansion, $2\pi\vec Q N$ (mod $2\pi$), smoothly and monotonically completes a full period in the process of describing the energy manifold. In fact, provided that $\vec Q$ does not satisfy any resonant condition~\cite{bazzani_analysis_1997}, $\vec p \cdot \vec Q = q$ ($\vec p \in \mathbb{Z}^3$, $q \in \mathbb{Z}$), the energy manifold of a particle can be recovered with the simple change of variable $\vec \Theta = 2\pi\vec Q N$. By doing so, one constructs an invariant torus, $\tilde \Psi$, directly from the time series of eq.~(\ref{eq:psi_x}). Denoting the projections of the torus in all three planes as $\tilde \Psi_x$, $\tilde \Psi_y$ and $\tilde \Psi_\zeta$, one can write:
\begin{equation}
\begin{aligned}
    \tilde \Psi^{(N_h)}_{x}(\vec \Theta) &= \tilde X- i  \tilde P_x = \sum_{k=0}^{N_h} {A_k\ e^{i[\vec n_k \cdot \vec \Theta] }}\\
    \tilde \Psi^{(N_h)}_{y}(\vec \Theta) &= \tilde Y- i  \tilde P_y = \sum_{k=0}^{N_h} {B_k\ e^{i[\vec m_k \cdot \vec \Theta] }}\\
    \tilde \Psi^{(N_h)}_{\zeta}(\vec \Theta) &= \tilde Z- i  \tilde P_\zeta = \sum_{k=0}^{N_h} {C_k\ e^{i[\vec \ell_k \cdot \vec \Theta] }}\\
\end{aligned}
\label{eq:Psi_topology}
\end{equation}


\noindent where capital letters are used to distinguish the single-particle coordinates ($\tilde x,\tilde p_x$, etc.) from the underlying topological object ($\tilde X,\tilde P_x$, etc.) on which they reside. Broadly speaking, an $n$-dimensional torus is a periodic structure constructed as the product of $n$ independent circles, or periodicities such as $\vec \Theta = (\Theta_x,\Theta_y,\Theta_\zeta)$~\cite{masoliver_integrability_2011}. The above result is therefore deeply linked to the existence of the normal form. Indeed, this particular phase space --- where the motion takes the form of independent rotations --- informs us that the energy manifolds of regular trajectories are topologically described by hyperdimensional tori. By going into the Courant-Snyder phase space (or even physical phase space), this claim is no less true and the topology is preserved, albeit deformed, leading to eq.~(\ref{eq:Psi_topology}). In the resonant case, the topology is different and needs to be appropriately described with regards to its corresponding fixed point. For example, the 5-th order resonance shown in Fig.~\ref{fig:Henon_phasor_expansion} requires to consider the 5-cycle of the map (thereby lifting the resonant condition) to get a new expansion in the form of eq.~(\ref{eq:psi_x}) and construct the energy manifold with eq.~(\ref{eq:Psi_topology}).\\

As mentioned, outside of the resonant condition, a given particle eventually covers the entire surface of its invariant torus as $\Theta_j = 2\pi Q_j N$, $j \in \{x,y,\zeta\}$, varies with arbitrarily large $N$ values, eventually covering $\Theta_j \in [0,2\pi]$ (mod $2\pi$) densely. This can be seen, for example, from the invariant curves (1-torus) of the Hénon map depicted in Fig.~\ref{fig:Henon_phasor_expansion}, obtained from eq.~(\ref{eq:Psi_topology}) by letting $\tilde \Psi_y = \tilde \Psi_\zeta = 0$. It should be clear --- at least visually --- that the invariant curves ($\tilde X,\tilde P_x$) are continuous closed loops, in contrast with the iterations of the map ($\tilde x,\tilde p_x$) which are obtained from a stroboscopic sampling (\textit{i.e.} $2\pi Q_x N$) of this underlying topological object. The spectral amplitudes $A_k,\ B_k,\ C_k$ and the integer vectors $\vec n_k,\ \vec m_k,\ \vec \ell_k$ in eq.~(\ref{eq:Psi_topology}) are the same as in eq.~(\ref{eq:psi_x}). As such, the torus obtained is a numerical approximation of the \textit{truly invariant} one ($N_h \to \infty$), obtained from --- and fully consistent with --- the quasiperiodic expansion of eq.~(\ref{eq:psi_x}).\\


For systems with more than 2 dimensions, visualizing the invariant tori requires additional care. In particular, non-linear coupling effects make the analysis difficult by introducing \textit{smearing}~\cite{belanger_topo_2024} in the single-particle trajectories when observed in either the physical phase space, $\vec x$, or the Courant-Snyder phase space, $\vec{\tilde x}$, as shown in the bottom plot of Fig.~\ref{fig:4D_torus}a for the 4D Hénon map (see Appendix~\ref{sec:Henon_map} for details of the map). That being said, eq.~(\ref{eq:Psi_topology}) offers a way to study and visualize the periodic structure of the underlying tori by letting one of the angles, say $\Theta_x$, vary along a closed loop while fixing the other angles, here $\Theta_y$. By doing so, the projections of the tori are obtained in both transverse planes; $\tilde \Psi_x(\Theta_x)$ in $(\tilde x,\tilde p_x)$ and $\tilde \Psi_y(\Theta_x)$ in $(\tilde y, \tilde p_y)$ for the 4D case, as illustrated in the top plot of Fig.~\ref{fig:4D_torus}a. When the single-particle trajectory is observed over many turns, both angles are densely explored following $\vec \Theta = 2\pi\vec Q N$ and the azimuthal deformation of the tori gives rise to the smearing observed in the bottom plot. Non-linear coupling can also be observed in Fig.~\ref{fig:4D_torus}a (more evidently for the high-amplitude particle) as the non-zero area of the projection observed in the $(\tilde y, \tilde p_y)$ plane. This is to say that $\tilde \Psi_y$ varies as a function of $\Theta_x$ --- the very definition of coupled motion. \\

More importantly, one can note that the sum of the areas enclosed by the loop projections $\tilde \Psi_x(\Theta_x)$ and $\tilde \Psi_y(\Theta_x)$ is one of the integrals of motion, conserved under the Hamiltonian flow. By numerically calculating this area following the procedure presented in the upcoming section, one can reconstruct the equivalent normal form tori --- circular tori of equal cross-sectional area --- as shown in Fig.~\ref{fig:4D_torus}b.

\subsection{Integrals of Motion \label{sec:integrals_of_motion}}
Embedded in the very nature of a symplectic transformation is the fact that the enclosed \textit{area} of any loop drawn in phase space is to be conserved~\cite{meiss_symplectic_1992}. For a $2n$-dimensional phase space, this area is the sum of the $n$ projections, on the different conjugate planes of the system, of the $2n$-dimensional loop. In particular, by choosing a loop lying on the energy manifold of a particle --- and moreover obtained from the evolution of a particular periodicity of the system, $\Theta_j$ --- one obtains the Poincaré integral invariant~\cite{meiss_symplectic_1992,bazzani_analysis_1997}, defined as :
\begin{equation}
    I_j = I_{jx} + I_{jy} + I_{j\zeta}=\frac{1}{2\pi}\oint_{\Theta_j}\Big(\tilde P_xd\tilde X + \tilde P_yd\tilde Y +\tilde P_\zeta d\tilde Z\Big)
    \label{eq:I_j}
\end{equation}

\noindent with $j\in \{x,y,\zeta \}$. Following the construction of the invariant tori with eq.~(\ref{eq:Psi_topology}), the above integral takes a purely geometric interpretation, already represented graphically in earlier plots. In Fig.~\ref{fig:Henon_phasor_expansion}, $I_x = I_{xx}$  is the area enclosed by the invariant curve of a particle in the $(\tilde x,\tilde p_x)$ phase space. In Fig.~\ref{fig:4D_torus}, $I_x = I_{xx} + I_{xy}$ is the sum of the areas (identified with a common color) enclosed by the projections of a 4-dimensional $\Theta_x$ loop in the  $(\tilde x,\tilde p_x)$ and $(\tilde y,\tilde p_y)$ conjugate planes. The contribution to the integral can either be positive or negative, depending on the orientation of the loop projection (positive if travelled clockwise), as indicated with the directional arrows in Fig.~\ref{fig:4D_torus}. In that example, the $I_{xy}$ projection --- which comes from non-linear coupling --- is negative in sign and varies along the torus. That being said, the sum $I_x = I_{xx} + I_{xy}$ is expected to be invariant. In fact, for a general torus described by eq.~(\ref{eq:Psi_topology}), the result of the integral from eq.~(\ref{eq:I_j}) can be shown to yield (see Appendix~\ref{sec:Derivation_I_x} for derivation):
\begin{equation}
\scalebox{0.89}{$\displaystyle 
I_j^{(N_h)} =   \frac{1}{2}\sum_{k=0}^{N_h}\left(n_{j_k}\left|A_k\right|^2 + m_{j_k}\left|B_k\right|^2 + \ell_{j_k}\left|C_k\right|^2\right)  +  \epsilon_j(\vec \Theta) 
$}
\label{eq:I_j_solution}
\end{equation}

\noindent which carries a remaining $\vec \Theta$ dependence. For a pure KAM torus --- that is, the energy manifold of a Hamiltonian system --- the  $\vec \Theta$ dependence observed in each projection is required to cancel out, which allows us to write:
\begin{equation}
    \epsilon_j (\vec \Theta) = \Delta_{jx}(\vec \Theta) + \Delta_{jy}(\vec \Theta) + \Delta_{j\zeta}(\vec \Theta) = 0\quad \forall\quad \vec \Theta
    \label{eq:eps_j}
\end{equation}

\noindent where the $\Delta$ functions describe the variation of each projection as a function of the non-integrated angles due to coupling effects, as discussed in the next section. However, in the numerical approximation of a KAM torus, it might be the case that the remainder $\epsilon_j (\vec \Theta)$ does not exactly cancel out, which will be discussed in section~\ref{sec:numerical_remainder}. It is important to mention that the remainder $\epsilon_j(\vec \Theta)$ being zero or not is not an artifact of the truncation itself, but rather the result of the torus being Hamiltonian or not.\\

This Poincaré integral, expressed in eq.~(\ref{eq:I_j}) in terms of the Courant-Snyder variables, can equally be computed in the physical phase space or the normal form phase space. Since $\Theta_x$, $\Theta_y$ and $\Theta_\zeta$ are 3 \textit{independent} periodicities of the torus, they describe 3 topologically independent loops, where none of the loops can be deformed continuously into each other or shrunk to zero~\cite{masoliver_integrability_2011}. Hence, the 3 integrals of motion obtained via eq.~(\ref{eq:I_j}) are independent actions of the system.

\subsubsection{Angle-dependant projections \label{sec:angle-dependant}}
\begin{figure}[t!]
    \centering
    \includegraphics[width=\linewidth]{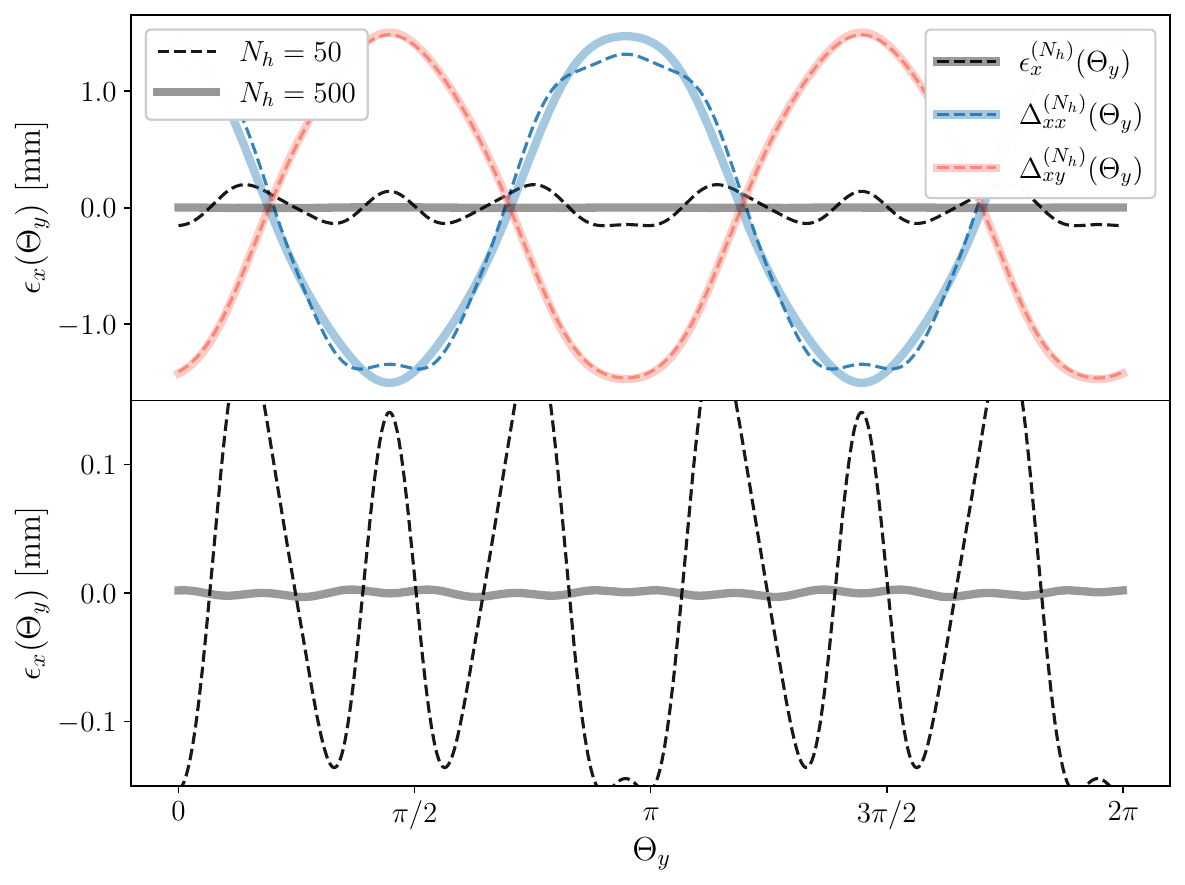}
    \caption{Evolution of the angle-dependant terms in the partial actions, $I_{xx}$ and $I_{xy}$, for the case of the 4D Hénon map (from Fig.~\ref{fig:4D_torus}a, outermost trajectory). Two truncation numbers, $N_h$, are shown. As $N_h \to \infty$, the sum of the angle-dependant terms goes to zero, $\epsilon_{x} = \Delta_{xx} + \Delta_{xy} \to  0$. With $N_h =50$ the variations along the torus are in the order of $0.1\%$ of the action.}
    \label{fig:Henon_action_evolution}
\end{figure}

As mentioned, the projections of a given loop in the conjugate planes of the system each carry an angular dependence, which only cancels out when the projections are summed together via eq.~(\ref{eq:I_j}), following eq.~(\ref{eq:eps_j}). This can be seen in the bottom plot of Fig.~\ref{fig:4D_torus}a, where the projection of the red loop in the $(\tilde y,\tilde p_y)$ plane is smaller than its green counterpart, evaluated at a different $\Theta_y$. In fact, for a loop around $\Theta_j$, the projections in the 3 different conjugate planes, $I_{jx}= \frac{1}{2\pi}\oint_{\Theta_j} \tilde P_x d\tilde X$, $I_{jy} = \frac{1}{2\pi}\oint_{\Theta_j} \tilde P_y d\tilde Y$ and $I_{j\zeta}= \frac{1}{2\pi}\oint_{\Theta_j} \tilde P_\zeta d\tilde Z$ take the form:
\begin{equation}
\left\{
\begin{aligned}
\quad I_{jx}^{(N_h)}(\vec \Theta) &= \frac{1}{2}\sum_{k=0}^{N_h} n_{j_k}\left|A_k\right|^2 + \Delta_{jx}(\vec \Theta)\\[1ex]
\quad I_{jy}^{(N_h)}(\vec \Theta) &= \frac{1}{2}\sum_{k=0}^{N_h} m_{j_k}\left|B_k\right|^2 + \Delta_{jy}(\vec \Theta)\\[1ex]
\quad I_{j\zeta}^{(N_h)}(\vec \Theta) &= \frac{1}{2}\sum_{k=0}^{N_h} \ell_{j_k}\left|C_k\right|^2 + \Delta_{j\zeta}(\vec \Theta)
\end{aligned}
\right.
\label{eq:I_projections}
\end{equation}

\noindent where for example:
\begin{equation}
\scalebox{0.90}{$\displaystyle 
\Delta^{(N_h)}_{xx}(\vec \Theta) = \frac{1}{2}\sum_{k=0}^{N_h}\sum_{l \ne k}^{N_h}{ \delta_{n_{x_k}}^{n_{x_l}}\Big(n_{x_k}|A_k||A_l|\cos(\varphi_{x_k}-\varphi_{x_l})\Big)}
$}
\label{eq:Delta_xx}
\end{equation}

\noindent in which $\delta_{a}^{b}$ is the Kronecker delta and $\varphi_{x_k} = n_{y_k}\Theta_y + n_{\zeta_k}\Theta_\zeta + \arg{[A_k]}$ is the parameter responsible for the dependence on $\Theta_y$ and $\Theta_\zeta$, the non-integrated angles. The different $\Delta$ functions are obtained from eq.~(\ref{eq:Delta_xx}) by permuting the plane considered, \textit{i.e.} where $\Delta_{xx} \to \Delta_{xy} \to \Delta_{x\zeta}$ is obtained by replacing $\vec n_k \to \vec m_k \to \vec \ell_k$ and $A_k \to B_k \to C_k$ and so on.\\

\noindent The evolution of these angle-dependant terms is shown in Fig.~\ref{fig:Henon_action_evolution} for the numerical approximation of a KAM torus obtained with different truncation numbers. The invariant torus considered is the one from the outermost particle from Fig.~\ref{fig:4D_torus}a. As one can see, the remainder $\epsilon_{x}(\vec \Theta) = \Delta_{xx}(\vec \Theta) + \Delta_{xy}(\vec \Theta)$ goes to zero as the number of harmonics, $N_h$, is increased.

\subsubsection{Numerical evaluation \label{sec:numerical_remainder}}
\begin{figure}[t!]
    \centering
    \includegraphics[width=\linewidth]{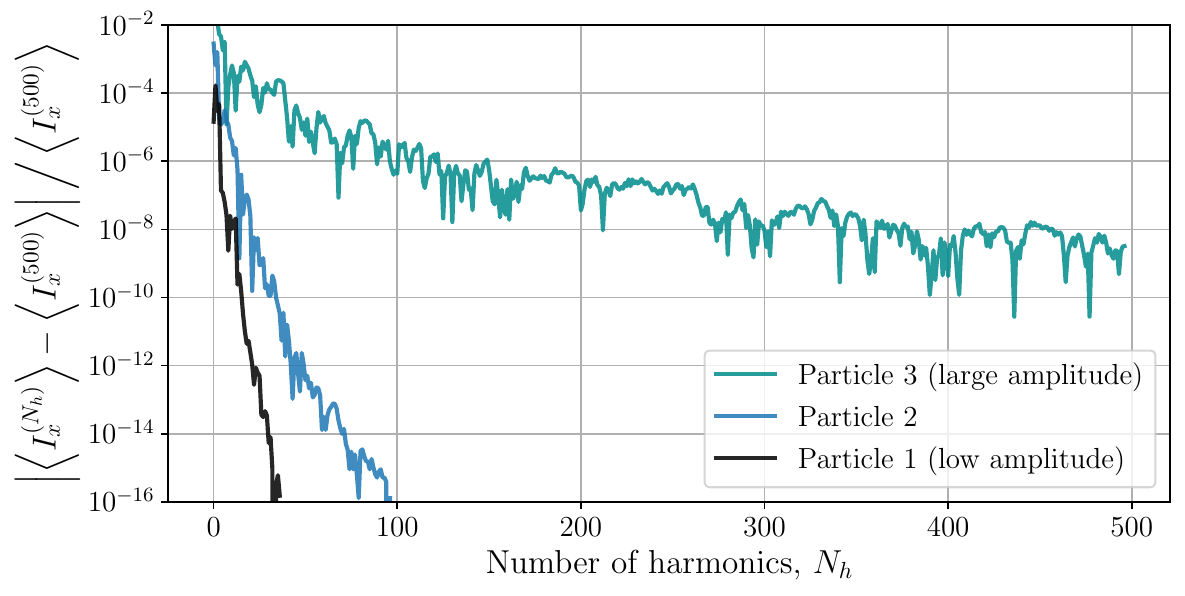}
    \caption{Relative error of the average action {\footnotesize$\ev{I_x^{(N_h)}}_{\vec \Theta}$}  as a function of the truncation number $N_h$ for the three particles of Fig.~\ref{fig:4D_torus}a. The convergence depends on the degree of non-linearity of the trajectory, which scales with the amplitude.}
    \label{fig:Henon_action_convergence}
\end{figure}

If the closed-form expansion provided by eq.~(\ref{eq:Psi_topology}) holds --- that is if the time series of eq.~(\ref{eq:psi_x}) accurately describes the single-particle motion and $\vec Q$ does not satisfy any resonant condition --- then all the aforementioned quantities can be computed numerically for the torus considered, which serves as an approximation to the energy manifold of the particle. However, as shown in Fig.~\ref{fig:Henon_action_evolution}, the angular dependence of this numerical description is not guaranteed to cancel out when summing up the different projections. In such cases, the action becomes difficult to evaluate. That being said, due to the presence of the cosine in eq.~(\ref{eq:Delta_xx}), the average over the angles --- for any $\Delta$ function --- always vanishes, \textit{i.e.} $\ev{\Delta}_{\vec \Theta}=0$. Moreover, since the amplitude of these oscillating functions converges towards zero as the number of harmonics is increased, the average value of $I_j^{(N_h)}(\vec \Theta)$ can be taken as an estimator of the true action for most truncation numbers. In fact, one can show that this average yields:
\begin{equation}
\scalebox{0.93}{$\displaystyle 
    \ev{I_j^{(N_h)}}_{\vec \Theta} = \frac{1}{2}\sum_{k=0}^{N_h}\left(n_{j_k}\left|A_k\right|^2 + m_{j_k}\left|B_k\right|^2 + \ell_{j_k}\left|C_k\right|^2\right)
$}
\label{eq:I_approximation}
\end{equation}

\noindent which is identical to eq.~(\ref{eq:I_j_solution}) with $\epsilon_j(\vec \Theta) = 0$, the action of a pure KAM torus. From there, one can asses the numerical validity of the approximation of a given torus by writing its actions as $I_j (\vec \Theta) = \ev{I_j}_{\vec \Theta} + \epsilon_j(\vec \Theta)$, separating the constant part from the angle-dependant part. Nevertheless, the estimation provided by eq.~(\ref{eq:I_approximation}) should be the final quoted quantity, since it converges much faster than the remainder, $\epsilon_j (\vec \Theta)$, when increasing the truncation number. As shown in Fig.~\ref{fig:Henon_action_convergence}, the convergence rate of this average depends on the amplitude of the particle or, more generally, on the degree of non-linearity of the lattice at the amplitude considered. The three particles taken here as examples are the ones of Fig.~\ref{fig:4D_torus}a, shown previously. For highly non-linear regimes, an infinite number of harmonics would, in principle, be required.

\section{Application to the LHC \label{sec:application_LHC}}

\begin{figure*}[ht!]
    \centering
    \subfloat{%
    \centering
    \hspace{-10mm}
    \includegraphics[width=1\textwidth,trim={0 2cm 0 0},clip]{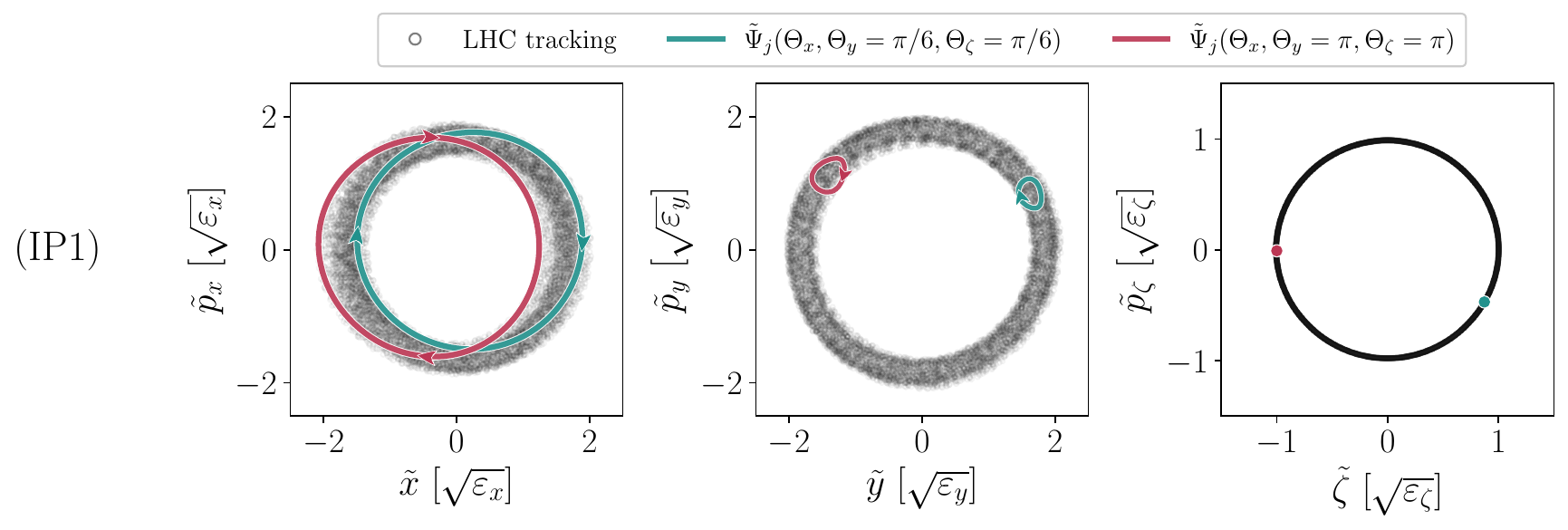}
    }
    
    \medskip 
    \centering
    \subfloat{%
    \centering
    \hspace{-10mm}
    \includegraphics[width=1\textwidth,trim={0 0 0 1.5cm},clip]{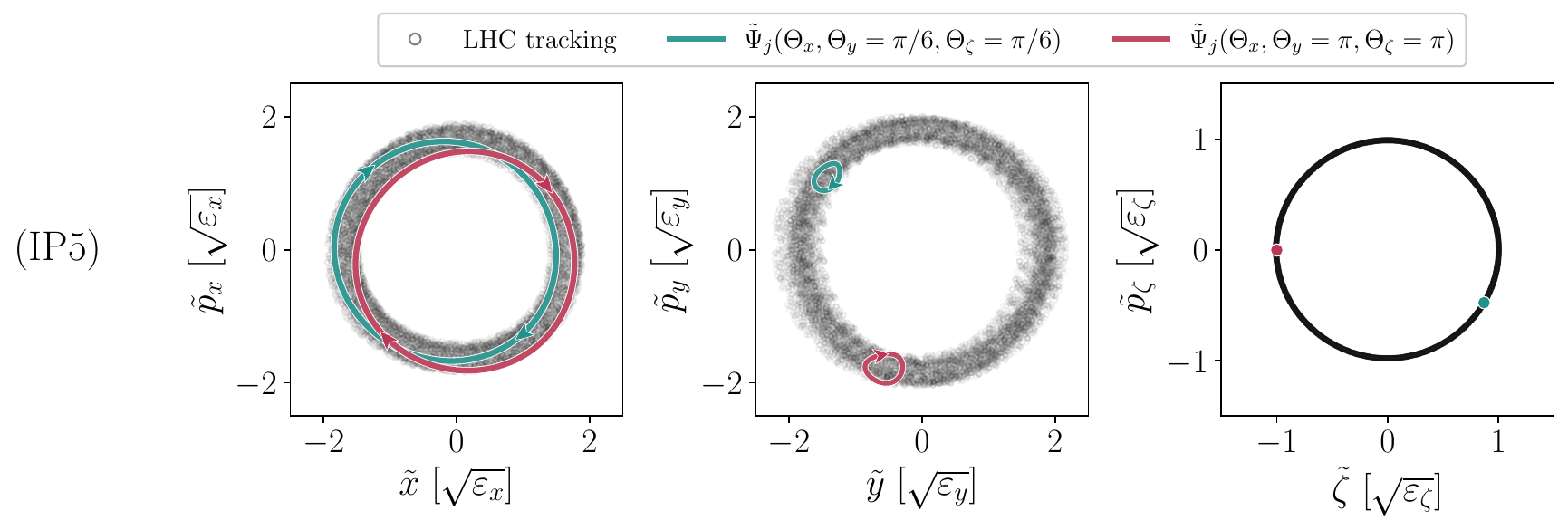}
    }
    \caption{Tracking results for a particle starting at $(x,y,\zeta) = (1.5 \sigma_x,2 \sigma_y,1 \sigma_\zeta)$ in the LHC. The same particle is observed separately in IP1 (top) and IP5 (bottom). The turn-by-turn data is shown, as well as the projections of the corresponding torus in all three planes, $\tilde \Psi_x$, $\tilde \Psi_y$, $\tilde \Psi_\zeta$, looping over $\Theta_x$. Exploring all the possible $\Theta_y$ and $\Theta_\zeta$ values allows to explain the important smearing of the trajectory. The horizontal action $I_x = I_{xx} + I_{xy} + I_{x\zeta}$ is obtained from the sum of the projected areas, identified with a common color.}
    \label{fig:LHC_example_ip1_ip5}
\end{figure*}

As a final demonstration of the framework introduced in this paper, the case of the 6-dimensional LHC at top energy is studied. The lattice considered is the one used as a baseline for the High-Luminosity upgrade of the LHC (HL-LHC), including beam-beam interactions in all the interaction points of the collider. The beam parameters considered can be found in Table~\ref{tab:HLLHC_parameters}.\\
\begin{table}[h!]
    \caption{HL-LHC parameters considered.}
    \centering
    \setlength{\tabcolsep}{3pt}
    \adjustbox{max width=\linewidth}{%
    \begin{tabular}{lccl}

        \toprule
        \textbf{Parameter}	            &                                    &	\textbf{Value} & \textbf{Unit} \\
        \midrule
      	
        Beam Energy 	                & $E$ 			& 7.0 & (TeV)\\
        Bunch intensity                 & $N_b$ 			& $1.13\times10^{11}$ & (p$^+$/b)\\
            Norm. Emittance             &   $\varepsilon_{x}^{_\text{N}},\varepsilon_{y}^{_\text{N}}$ & 2.5 & ($\upmu$m$\cdot$rad)\\
            Long. Emittance$^\dagger$   &   $\varepsilon_{\zeta}^{_\text{N}}$ & 62 & (mm$\cdot$rad)\\
            
            Beta at the IP		        & $\beta^*$		& 	15 & (cm)\\
        Half-crossing	                & $\theta_c/2$	&  $250$ & ($\upmu$rad) \\
        Crabbing angle	                & $\theta_{cc}$	&  $-190$ & ($\upmu$rad) \\
            Octupoles                   & $I_\text{oct}$ & $-60$ & (A)\\ 
            Tunes                       & $(Q_x,Q_y)_\text{c.o.}$   & $(62.312,60.319)$ & \\
                                        & $Q_{\zeta_\text{c.o.}}$   & $-0.002$ & \\
            Chromaticities              & $\Delta Q_{x_\text{c.o.}}$,$\Delta Q_{y_\text{c.o.}}$ & 15 &\\
            
        \bottomrule\vspace{-2mm}\\
        \multicolumn{2}{l}{$^\dagger$ See Appendix~\ref{sec:xsuite_coordinates} for details.}
    \end{tabular}}
    \label{tab:HLLHC_parameters}
 \end{table}

Alongside electron cloud effects, beam-beam effects are the main source of non-linearities in the LHC, leading to important beam losses and the reduction of dynamic aperture by several sigmas~\cite{herr_beam-beam_2014-1,belanger_bunch-by-bunch_2024-2}. Due to the complex nature of these interactions, their effect on the dynamics of the beam is typically studied in the transverse plane only, after resorting to several approximations. However, using the Hirata method~\cite{hirata_symplectic_2003,cern_xsuite_nodate}, beam-beam effects can be accurately modelled in 6D tracking codes, including their impact on the longitudinal plane.\\

For this complex 6D case, synchro-betatron coupling effects arise and the dynamics becomes difficult to analyze. However, the system is still Hamiltonian and the formalism introduced above, mainly driven by eq.~(\ref{eq:psi_x}) and eq.~(\ref{eq:Psi_topology}), can still be used. Using NAFF to extract $N_h=100$ harmonics in all three planes, the approximate tori can be constructed to evaluate the action and understand the heavily smeared single-particle trajectories observed in the tracking results. Such an example is shown in Fig.~\ref{fig:LHC_example_ip1_ip5}, where an off-momentum particle starting from $(x,y,\zeta) = (1.5 \sigma_x,2 \sigma_y,1 \sigma_\zeta)$ is tracked over $10^5$ turns. Slices of the corresponding torus (looping over $\Theta_x$ as an example) are shown on top of tracking data for different values of $\Theta_y$ and $\Theta_\zeta$. One can see that the turn-by-turn data can be obtained from a stroboscopic sampling ($\vec \Theta = 2\pi\vec Q N$) of this torus, turn after turn. In fact, by exploring all of the $\vec \Theta$ angle values, one can show that the entirety of the smeared region is eventually covered. Additionally, the horizontal action $I_x = I_{xx} + I_{xy} + I_{x\zeta}$ is obtained from the sum of the projected areas of the $\Theta_x$ loop in all three planes, identified with a common color. To show --- at least visually --- the effect of symplectic transformations along the machine, two independent monitors are installed, one in IP1 and one in IP5, for comparison.\\

\begin{figure*}[ht!]
    \centering
    \subfloat{%
    \centering
    \hspace{-10mm}
    \includegraphics[width=1\textwidth,trim={0 1.45cm 0 0},clip]{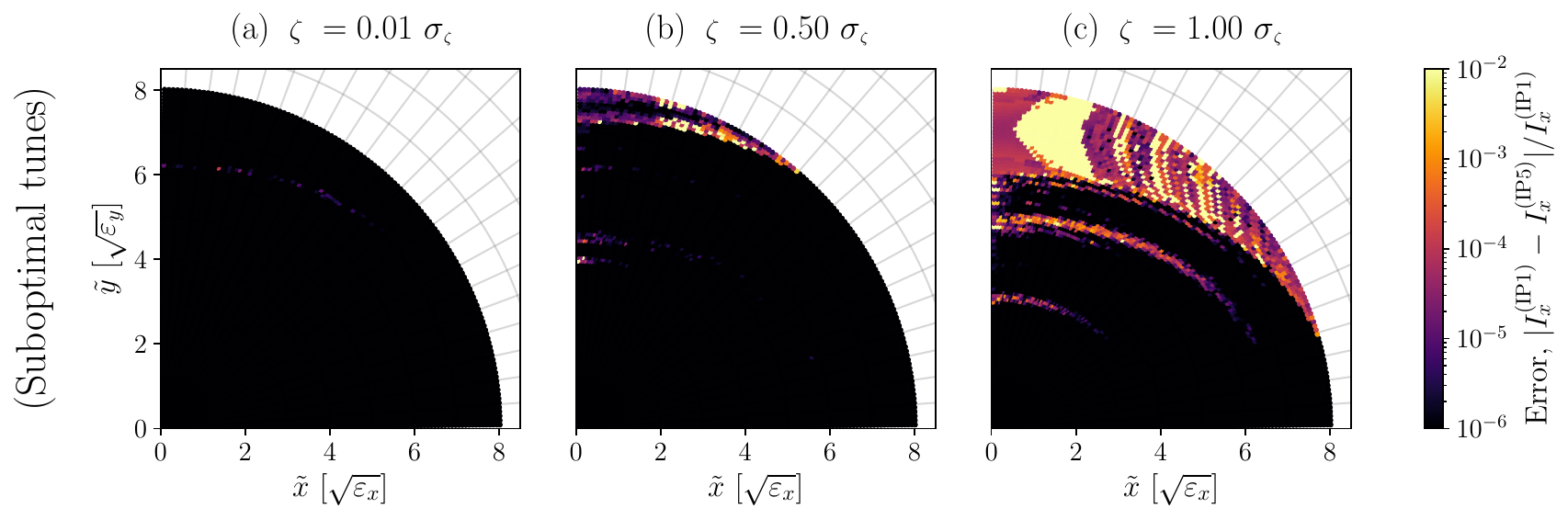}
    }
    
    \medskip\vspace{-2mm} 
    \centering
    \subfloat{%
    \centering
    \hspace{-10mm}
    \includegraphics[width=1\textwidth,trim={0 0 0 1.0cm},clip]{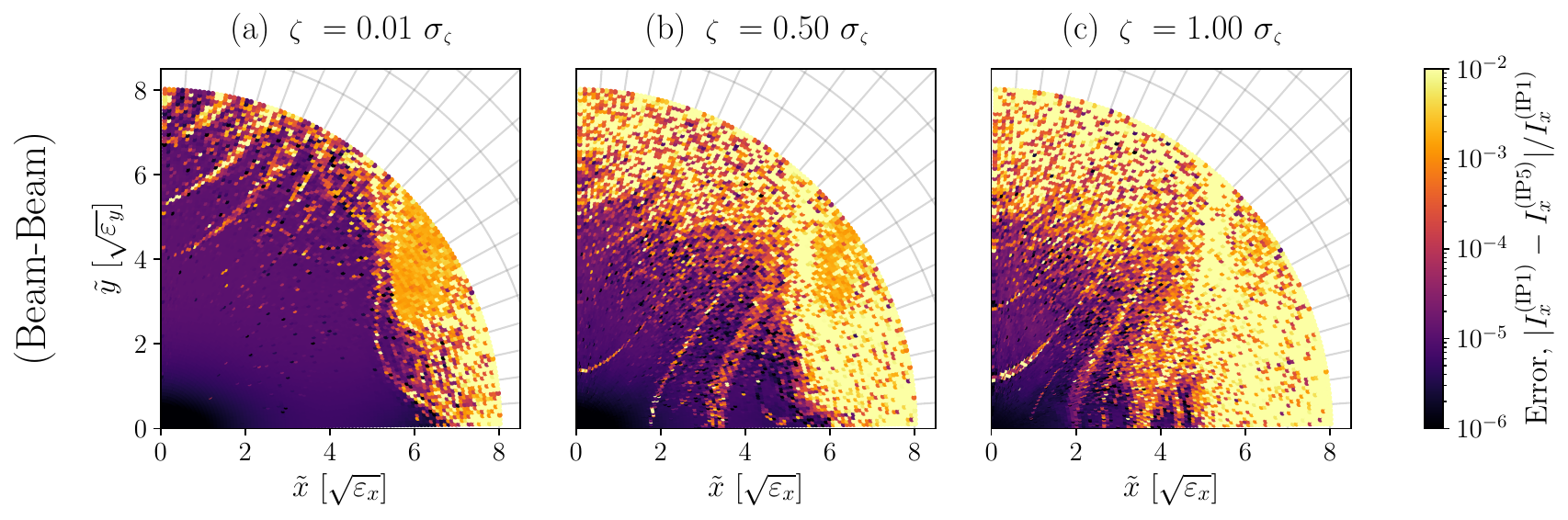}
    }
    \caption{Evaluation of the variation of the horizontal action, $I_x$, between IP1 and IP5 for two different versions of the HL-LHC lattice from Table~\ref{tab:HLLHC_parameters}. (Top) A suboptimally tuned $(Q_x,Q_y)_\text{c.o.} = (62.315 , 60.331)$ lattice without beam-beam interactions. (Bottom) The baseline HL-LHC lattice at top energy with beam-beam interactions in all IPs. Three initial off-momentum coordinates are sampled ($\zeta = \{0.01 \sigma_\zeta,0.50 \sigma_\zeta,1.00 \sigma_\zeta\}$) and a dense grid of transverse initial positions is considered in each case. The presence of beam-beam interactions increases significantly the number of resonances. The trajectory of a particle taken from the rightmost bottom plot is shown in Fig.~\ref{fig:LHC_example_ip1_ip5}.}
    \label{fig:LHC_Ix_error}
\end{figure*}

The integrals of motion shall respect the defining feature of being invariant in time --- turn after turn --- as well as along the machine --- in $s$. The latter form of conservation is easier to evaluate numerically since it trades computing time for memory. As such, a similar study to the one shown graphically in Fig.~\ref{fig:LHC_example_ip1_ip5} was repeated for a dense grid of particles in the 6-dimensional phase space and the variation of the horizontal action, $I_x$, between IP1 and IP5 was evaluated. The results of this study are shown in Fig.~\ref{fig:LHC_Ix_error} for two different machines based on Table~\ref{tab:HLLHC_parameters}: a suboptimally tuned $(Q_x,Q_y)_\text{c.o.} = (62.315 , 60.331)$ version of the HL-LHC baseline lattice without beam-beam (top); as well as the full HL-LHC baseline lattice including beam-beam interactions (bottom).  The particles were tracked for $10^5$ turns and $N_h=100$ harmonics were extracted. It is noteworthy that a smaller number of turns could have been considered for the numerical evaluation of a simpler lattice. However, in presence of beam-beam effects, many spectral lines are introduced, which requires a higher number of turns to accurately resolve the spectrum. As an example of a simpler system, the reader is referred to the case of the 2D Hénon map, shown in Appendix~\ref{sec:henon_error}.\\

The suboptimally tuned case (top plots, Fig.~\ref{fig:LHC_Ix_error}) shows that the integrals of motion, evaluated numerically on the basis of eq.~(\ref{eq:I_approximation}), are indeed well-preserved between the IPs. In these plots, resonance conditions appear as bright stripes where the integrals of motions cannot be properly evaluated due to the breaking of the topological condition required for eq.~(\ref{eq:Psi_topology}) to hold. In the bottom plots, one can see that the addition of beam-beam interactions increases significantly the number of resonances and makes the evaluation of the integrals of motion more challenging. In principle, the variation of the action between the IPs could be reduced for a large region of the phase space (purple region, $\sim 10^{-5}$ action error) by increasing the number of turns or the number of harmonics considered due to the high degree of non-linearities present in the lattice. Indeed, as shown in Fig.~\ref{fig:LHC_example_ip1_ip5} (particle taken directly from the bottom plot of Fig.~\ref{fig:LHC_Ix_error}c), the motion in that region remains regular and is well-described by the approximate invariant torus extracted from the tracking. On the contrary, for resonant cases, such a behaviour cannot be recovered and the picture is completely different. However, the labelling of resonant cases is a subtle issue which should not be trivialized. The reader is referred to Appendix~\ref{sec:continuity} for additional details. As a final note, it appears that the continuity of the action is an important indicator to consider when trying to determine whether a given particle lies on a KAM torus or not, as highlighted in Appendix~\ref{sec:henon_error}.



\section{Summary}
This paper showed that invariant tori can be constructed numerically to approximately describe the energy manifold of single-particle trajectories based on tracking data. From there, the integrals of motion can be found based on the area of the tori projections in all conjugate planes --- in 2D, 4D or 6D --- and computed numerically. The formalism described is based on the KAM theorem, which guarantees the existence of some regular trajectories expressed in terms of complex Fourier series. After extracting the harmonics of said series from tracking data, it is shown that the underlying topological object --- an invariant torus --- can be constructed. This conceptualization of the motion allows to describe coupled motion (transverse and synchro-betatron coupling) in a natural manner and accounts for the heavily smeared trajectories observed in phase space. Examples are shown based on the 2D and 4D Hénon map. As a final numerical demonstration, it is shown that the HL-LHC baseline lattice can be studied with this formalism, including beam-beam effects in 6D. This pragmatic and descriptive approach appears promising to further the understanding of non-linear motion in particle accelerators, beyond the reach of analytic calculations. In particular, the closed-form expression presented to describe invariant tori can be used to quantify non-linearities and tori deformation for compensation studies, as well as non-linear matching problems. That being said, some important challenges remain with regard to the transition from regular to resonant to chaotic motion, which was not explored in this paper.

\begin{acknowledgments}
The authors would like to thank R.~Baartman, T.~Planche and D.~Kaltchev for their help and guidance throughout the redaction of this paper. Moreover, the important contribution of K.~Paraschou, through numerous discussions and arguments, needs to be highlighted. Finally, special thanks to C.E.~Montanari, M.~Giovannozzi and A.~Bazzani for their feedback on the work.
\end{acknowledgments}


{
\noindent\begin{center}
    \rule{0.9\linewidth}{0.5pt}\vspace{-3.95mm}
    \rule{0.75\linewidth}{1pt}\vspace{-3.95mm}
    \rule{0.5\linewidth}{1.5pt}
\end{center}
}

\newpage
\appendix
\section{Single-particle tracking\label{sec:xsuite_coordinates}}

The formalism developed in this paper is intended to be general and independent of the tracking code considered. That being said, it is worth mentioning that the notation used throughout the paper is based on the Xsuite code~\cite{Iadarola:2023fuk}, developed at CERN. Xsuite is a 6D single particle symplectic tracking code used to compute the trajectories of individual relativistic charged particles in circular accelerators. Following standard practices, the coordinates presented in Section~\ref{sec:coordinates} correspond to deviations from the reference trajectory of a particle with speed $\beta_0c$ and momentum $P_0$. The transverse canonical momenta $P_x$ and $P_y$ are normalized following:
\begin{equation}
    p_x = P_x/P_0\quad \text{and}\quad p_y = P_y/P_0\ ,
\end{equation}

\noindent yielding the unitless momentum considered in this paper. In the longitudinal plane, the longitudinal deviation from the reference particle, as well as the energy deviation, are considered. The coordinates $(\zeta,p_\zeta)$ therefore read:
\begin{align}
   \zeta = s - \beta_0 ct \quad \text{,}\quad p_\zeta = \frac{1}{\beta_0^2}\frac{E-E_0}{E_0}\ .
    \label{eq:zeta_pzeta}
\end{align}

\begin{figure}[h!]
    \centering
    \includegraphics[width=1\linewidth]{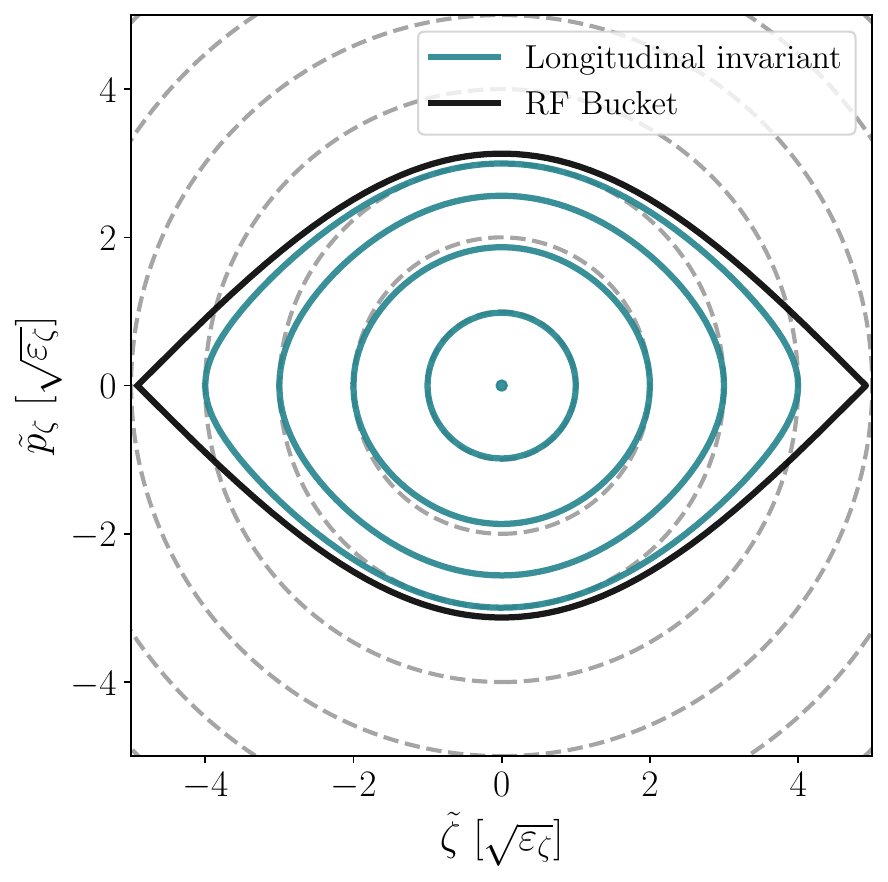}
    \caption{Longitudinal phase space of the LHC for the Courant-Snyder coordinates $(\tilde \zeta,\tilde p_\zeta)$. The invariant curves of the Hamiltonian are shown in blue, and the separatrix (RF Bucket) is shown in black. The grey dashed lines correspond to multiples of $\sigma_\zeta$, as seen in the Courant-Snyder phase space.}
    \label{fig:RF_Bucket}
\end{figure}

\noindent Just like the transverse coordinates, $(\zeta,p_\zeta)$ can be transformed into the Courant-Snyder phase space to obtain $(\tilde \zeta,\tilde p_\zeta)$ with the help of the $W$-matrix, as discussed in Section~\ref{sec:coordinates}. By doing so, the longitudinal motion lies on circles for small energy deviations and is ultimately limited by the RF bucket for large energy deviations, as shown in Fig.~\ref{fig:RF_Bucket}, which is given as an example of the metric. In the paper, the longitudinal coordinates are often given in units of $\sigma_\zeta$, which coincide with the dashed lines of Fig.~\ref{fig:RF_Bucket}, in the Courant-Snyder phase space.\\

In Table~\ref{tab:HLLHC_parameters}, the normalized longitudinal emittance is given in units of mm$\cdot$rad, which requires clarifications. The design bucket area of the LHC is typically given as 7.9 eVs with a bunch area of 2.5 eVs, considering $2\sigma$ in both dimensions of the $(\Delta t,\Delta E)$-space. Dividing by $4\pi$, the corresponding emittance is therefore of $\varepsilon_{_{\Delta t}} = 0.2$~eVs. To convert this emittance to the proper units for the $(\zeta,p_\zeta)$ phase space --- by analogy to the transverse emittance --- one needs to multiply by $\beta_0 c$ and divide by $\beta_0^2 E$ as per eq.~(\ref{eq:zeta_pzeta}) above. Hence, normalized to $\beta_0 \gamma_0$, the design value of $\varepsilon^\text{N}_\zeta$ is $\varepsilon^\text{N}_\zeta = \frac{\beta_0 c}{\beta_0^2 E_0}\varepsilon_{_{\Delta t}} (\beta_0 \gamma_0) = 64$~mm$\cdot$rad at 7~TeV.\\




For additional details on the general Hamiltonian and the physics included in the Xsuite code, the reader is referred to the Xsuite manual~\cite{cern_xsuite_nodate}.

\noindent 
\newpage
\section{Hénon Map \label{sec:Henon_map}}
The Hénon map~\cite{Henon_numerical_1969} is one of the simplest model of non-integrable Hamiltonian maps which exhibits chaotic behaviour. It is an ideal mathematical tool to discuss the features of non-linear dynamics with reproducible results. Moreover, the map can be constructed from a sequence of thin magnetic elements, as discussed below.

\subsection{2D case}

Let's consider the 2D case of a thin sextupole followed by a linear segment with phase advance $\mu_x$ and beta function $\beta_x$ at its entry and exit. Using the complex notation $\tilde \psi_x = \tilde x-i\tilde p_x$, one can show that the one-turn map of this system can be written in closed-form in the Courant-Snyder phase space. The application of the resulting map, $\mathcal{M}_\text{H}$, on a given complex point $\tilde \psi_x$ yields the new point:
\begin{equation}
    \mathcal{M}^{(2\text{D})}_\text{H}\ \tilde \psi_x = e^{i[2\pi\mu_x]}\Big[\tilde \psi_x - i\frac{a}{4}(\tilde \psi_x^* + \tilde \psi_x)^2\Big]
    \label{eq:Henon2D}
\end{equation}

\noindent where $a=-\frac{1}{2}k_2\beta_x^{3/2}$ is the strength of the non-linearity as a function of the sextupole strength, $k_2$, and the beta function at the entrance of the linear segment. The particular case where $a=1$ corresponds to the well studied Hénon map (which can also be obtained via a non-symplectic rescaling $\tilde \psi_x\to \tilde \psi_x/a$)~\cite{bazzani_normal_1994-2}.

\subsection{4D case}
By extension to the previous case, the application of the very same transformations on a set of 4D coordinates $\tilde \psi_x = \tilde x  - i\tilde p_x$ and $\tilde \psi_y = \tilde y  - i\tilde p_y$ yields the new points:

\begin{align}
    \left\{\begin{aligned}
         \mathcal{M}^{(4\text{D})}_\text{H}\ \tilde \psi_x &= e^{i\mu_x}\bigg[\tilde \psi_x - i\frac{a}{4}\Big((\tilde \psi_x^* + \tilde \psi_x)^2 - \rho (\tilde \psi_y^* + \tilde \psi_y)^2\Big)\bigg]\\
         \mathcal{M}^{(4\text{D})}_\text{H}\ \tilde \psi_y &=  e^{i\mu_y}\bigg[\tilde \psi_y + i\frac{a\cdot \rho }{2}(\tilde \psi_x^* + \tilde \psi_x)(\tilde \psi_y^* + \tilde \psi_y)\bigg]
    \end{aligned}\right.
\end{align}

\noindent where $\rho = \beta_y/\beta_x$ and $a=-\frac{1}{2}k_2\beta_x^{3/2}$ once again. The case where $a=1$ corresponds to the well-studied 4D Hénon map~\cite{bazzani_normal_1994-2}. The $\rho$ parameter is responsible for the non-linear coupling between the two planes. One can see that for $\rho=0$ or $\tilde \psi_y=0$, the 2D Hénon map is recovered in the horizontal plane for $\tilde \psi_x$.

\section{NAFF, summary of the algorithm\label{sec:NAFF}}
Let's consider a quasiperiodic function $\psi(N)$ known over a finite time span $N \in [0,T]$. The spectral amplitude $A(\nu)$ can be computed as a function of $\nu$ (assuming continuous $\nu$, equivalent to zero-padding FFT) following:
\begin{equation}
    A(\nu) = \frac{1}{T}\sum_{N=0}^{T-1} \psi(N) \cdot e^{i[2\pi \nu \cdot N]}\cdot \chi_p(N)
    \label{eq:DFT}
\end{equation}
where the Hann window $\chi_p(N)$, of order $p$ and centered on the dataset, is given by:
\begin{equation}
    \chi_p(N) = \frac{2^p (p!)^2}{(2p)!}\left(1+\cos(\frac{2\pi (N- T/2)}{T})\right)^p
    \label{eq:hanning}
\end{equation}

From there, the main frequency $\nu_0$ can be found by maximizing eq.~(\ref{eq:DFT}) using, for example, a Newton–Raphson method. With the windowing function of eq.~(\ref{eq:hanning}), $\nu_0$ can be found with an accuracy that scales with $1/T^{2p+2}$, as compared to $1/T$ for a simple FFT~\cite{simo_introduction_1999}. The key aspect of Laskar's approach however, is to then subtract the contribution of $\nu_0$ from the original signal, and repeat the procedure to obtain the second dominant frequency, $\nu_1$, and so on.

\section{Derivation of the integrals of motion \label{sec:Derivation_I_x}}

The Poincaré integral invariant given by eq.~(\ref{eq:I_j}) can be separated into its projections onto the different canonical planes as discussed in section~\ref{sec:integrals_of_motion}. We are therefore looking to compute one of these projections, \textit{e.g.} $I_{xx}$, given by:
\begin{equation}
    I_{xx}= \frac{1}{2\pi}\oint_{\Theta_x} \tilde P_x d\tilde X = \frac{1}{2\pi}\int_{0}^{2\pi}{\left(\tilde P_x \frac{\partial \tilde X}{\partial \Theta_x} \right)d\Theta_x}
\end{equation}

\noindent Starting from eq.~(\ref{eq:Psi_topology}), one can assume a set of coordinates, $(\tilde X,\tilde P_x)$, written in the complex form $\tilde \Psi_x = \tilde X - i\tilde P_x$ following: 
\begin{equation}
    \tilde \Psi^{(N_h)}_{x}(\vec \Theta) = \tilde X- i  \tilde P_x = \sum_{k=0}^{N_h} {A_k\ e^{i[\vec n_k \cdot \vec \Theta] }}
    \label{eq:annex_Psix}
\end{equation}

\noindent For any fixed $\Theta_y$ and $\Theta_\zeta$, the loop $\tilde \Psi_x(\Theta_x)$ thereby formed lies on the torus given by eq.~(\ref{eq:annex_Psix}) --- as shown graphically in Fig.~\ref{fig:4D_torus} --- for which we wish to calculate the area. To facilitate the integration over $\Theta_x$, one can introduce the phase parameter $\varphi_{x_k} = n_{y_k}\Theta_y + n_{\zeta_k}\Theta_\zeta + \arg{[A_k]}$ such that:
\begin{equation}
    \tilde \Psi^{(N_h)}_{x}(\Theta_x) = \sum_{k=0}^{N_h} {\left|A_k\right|\ e^{i[n_{x_k}\Theta_x + \varphi_{x_k}] }}
    \label{eq:annex_Psix_parameter}
\end{equation}

\noindent Taking the real part of eq.~(\ref{eq:annex_Psix_parameter}) for the position and the imaginary part for the momentum, one can directly write:
\begin{align}
    \tilde P_x \frac{\partial \tilde X}{\partial \Theta_x} = &\sum_{k=0}^{N_h}\left(-|A_k|\right)\sin{[n_{x_k} \Theta_x + \varphi_{x_k}] } \nonumber \\
    &\times \sum_{l=0}^{N_h}\left(-n_{x_l}|A_l| \right)\sin{[n_{x_l} \Theta_x + \varphi_{x_l}] }
\end{align}

\noindent To proceed with the integral, the following orthogonal identity can be used:
\begin{align}
    \int_0^{2\pi}\sin(n_{x_k}\Theta_x + \varphi_{x_k}) & \sin(n_{x_l}\Theta_x + \varphi_{x_l}) d\Theta_x \\
    &= \pi \cos(\varphi_{x_k} - \varphi_{x_l})\cdot \delta_{n_{x_k}}^{n_{x_l}} \nonumber
\end{align}

\noindent where $\delta_{n_{x_k}}^{n_{x_l}} = 1$ if $n_{x_k} = n_{x_l}$ and $0$ otherwise. After integration, we therefore end up with:
\begin{equation}
\scalebox{0.95}{$\displaystyle 
    I_{xx} = \frac{1}{2}\sum_{k=0}^{N_h}\sum_{l =0 }^{N_h}{ \delta_{n_{x_k}}^{n_{x_l}}\Big(n_{x_k}|A_k||A_l|\cos(\varphi_{x_k}-\varphi_{x_l})\Big)}  
$}
\end{equation}

\noindent where the summation is taken over the various spectral lines of the horizontal spectrum, $\tilde \Psi_x$. One can notice that the result is tied to the phase parameter $\varphi$, which depends on the non-integrated angles $\Theta_y$ and $\Theta_\zeta$. Hence, it is relevant to separate the terms that are angle-independent from the others (\textit{i.e.} when $l=k$, $\varphi_{x_k}-\varphi_{x_l} = 0$) to expose the dependence on $(\Theta_y,\Theta_\zeta)$, yielding:
\begin{equation}
    I_{xx}(\vec \Theta) = \frac{1}{2}\sum_{k=0}^{N_h} n_{x_k}\left|A_k\right|^2 + \Delta_{xx}(\vec \Theta)
\end{equation}
where the angle dependant part is explicitly given by:
\begin{equation}
\scalebox{0.95}{$\displaystyle 
\Delta^{(N_h)}_{xx} = \frac{1}{2}\sum_{k=0}^{N_h}\sum_{l \ne k}^{N_h}{ \delta_{n_{x_k}}^{n_{x_l}}\Big(n_{x_k}|A_k||A_l|\cos(\varphi_{x_k}-\varphi_{x_l})\Big)}
$}
\label{eq_an:Delta_xx}
\end{equation}

\noindent Generalizing, the Poincaré integral invariant $I_j = I_{jx} + I_{jy} + I_{j\zeta}$ (for $j\in \{x,y,\zeta\}$) can be obtained by summing the three projections:
\begin{equation}
\left\{
\begin{aligned}
\quad I_{jx}^{(N_h)}(\vec \Theta) &= \frac{1}{2}\sum_{k=0}^{N_h} n_{j_k}\left|A_k\right|^2 + \Delta_{jx}(\vec \Theta)\\[1ex]
\quad I_{jy}^{(N_h)}(\vec \Theta) &= \frac{1}{2}\sum_{k=0}^{N_h} m_{j_k}\left|B_k\right|^2 + \Delta_{jy}(\vec \Theta)\\[1ex]
\quad I_{j\zeta}^{(N_h)}(\vec \Theta) &= \frac{1}{2}\sum_{k=0}^{N_h} \ell_{j_k}\left|C_k\right|^2 + \Delta_{j\zeta}(\vec \Theta)
\end{aligned}
\right.
\end{equation}

\noindent where the different $\Delta$ functions are obtained from eq.~(\ref{eq_an:Delta_xx}) by permuting the plane considered, \textit{i.e.} where $\Delta_{xx} \to \Delta_{xy} \to \Delta_{x\zeta}$ is obtained by replacing $\vec n_k \to \vec m_k \to \vec \ell_k$ and $A_k \to B_k \to C_k$ and so on.

\section{Continuity \& Uniqueness \label{sec:continuity}}
\begin{figure}[t!]
    \centering
    \includegraphics[width=1\linewidth]{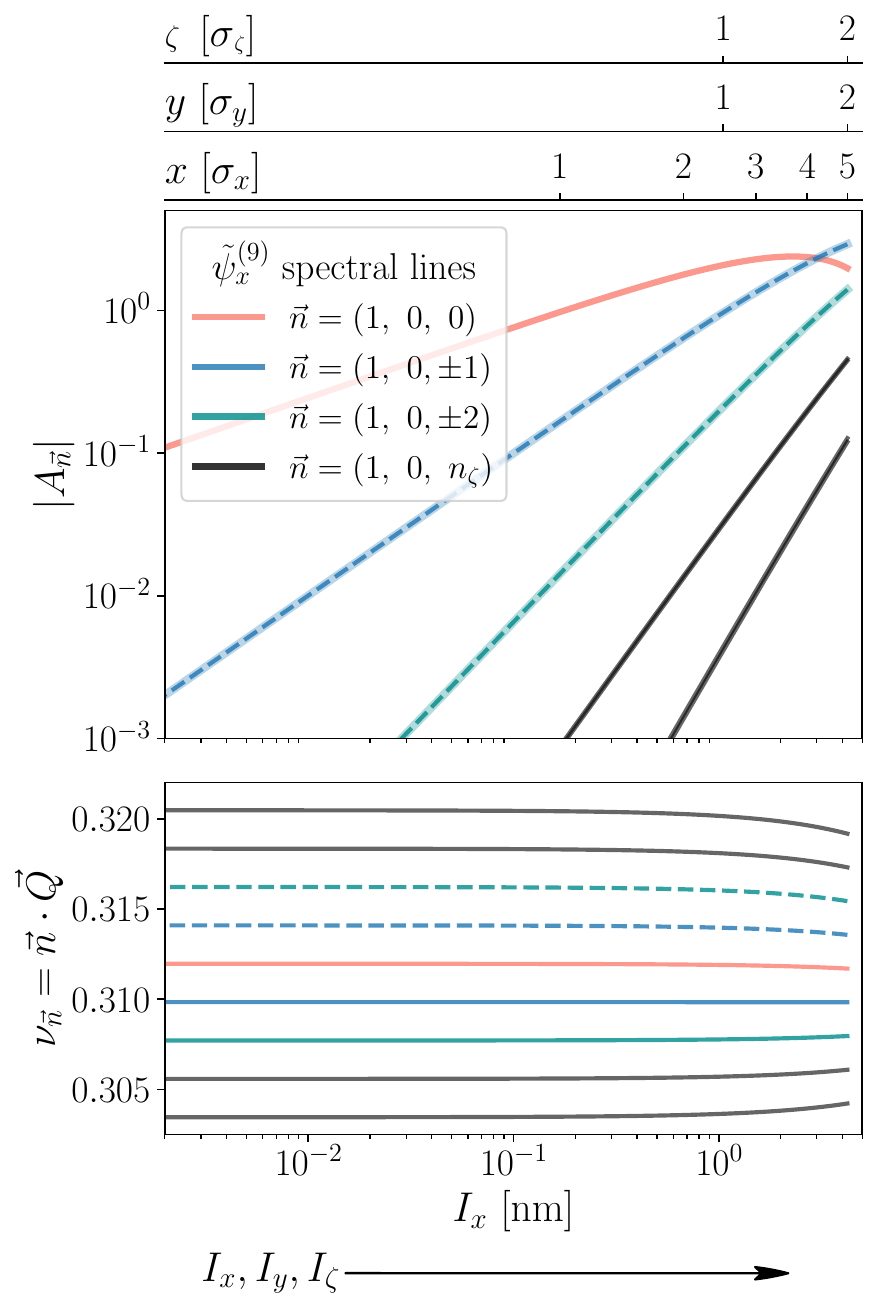}
    \caption{Continuity of the main line and the first few synchrotron sidebands from the horizontal spectrum, $\tilde \psi_x^{(9)}$, for a particle moving from the origin, $(x,y,\zeta) = (0,0,0)$ to a high amplitude point $(x,y,\zeta) = (5 \sigma_x,2 \sigma_y,2 \sigma_\zeta)$ in the LHC (see Table~\ref{tab:HLLHC_parameters} for details, without beam-beam interactions). The spectral amplitudes $A_{\vec n}$, the fundamental frequencies $\vec Q$ and the harmonics $\vec n\cdot \vec Q$ are all smooth functions of the action, $I_x$. $I_y$ and $I_\zeta$ are also increased as the particle is moved towards the high amplitude point. Results obtained via tracking.}
    \label{fig:LHC_continuity}
\end{figure}
It can be shown that a given family of neighbouring KAM trajectories (a KAM region) smoothly depends on the actions, far away from resonances~\cite{simo_introduction_1999}. In fact, the Fourier expansion coefficients in each plane, as well as the single-particle fundamental frequencies, are expected to be continuous and differentiable, such that:
\begin{equation}
    \frac{\partial A_{\vec n}}{\partial I_j},
    \frac{\partial B_{\vec m}}{\partial I_j},
    \frac{\partial C_{\vec \ell}}{\partial I_j}
    \ \text{and}\
    \frac{\partial \vec Q}{\partial I_j}
    \quad 
    \text{are smooth}
    \label{eq:continuity}
\end{equation}

\noindent for $j \in \{x,y,\zeta\}$, where the spectral lines need to be indexed according to their corresponding integer vector ($\vec n$, $\vec m$ or $\vec \ell$) to be unambiguously identified, instead of sorting them by decreasing amplitude. An important consequence of eq.~(\ref{eq:continuity}), and in particular of the smooth evolution of $\vec Q$, is that if the betatron frequencies (the \textit{tunes}) are well-defined on the closed-orbit of the machine (which is generally the case), then so are the particles fundamental frequencies in the neighbouring KAM region around the origin. To illustrate this claim, the continuity of some spectral lines (as a function of $I_x$) from the horizontal spectrum of the LHC is shown in Fig.~\ref{fig:LHC_continuity}. It can be seen that although the first synchrotron sidebands overtake the main line in terms of amplitude around $\zeta \approx 1.5\sigma_\zeta$, the continuity condition of eq.~(\ref{eq:continuity}) allows us to unambiguously identify the main line and its frequency (the single-particle fundamental frequency, $Q_x$) as being the one in continuity with the closed-orbit betatron frequency when $(x,y,\zeta)\to (0,0,0)$.\\

The labelling, or mislabelling, of the spectral lines is a subtle issue which requires careful consideration and is not to be trivialized. It is commonly assumed that the frequency of the highest spectral line, $\nu_0$, corresponds to the fundamental frequency of the invariant torus, $\nu_0 \leftrightarrow [\nu_{(1,0,0)} = Q_x]$, which is clearly not the case in the example provided in Fig.~\ref{fig:LHC_continuity} for high amplitude particles. This leads to mislabelling issues which seemingly breaks the continuity expected from eq.~(\ref{eq:continuity}). In fact, inspecting eq.~(\ref{eq:psi_x}) together with eq.~(\ref{eq:nu_k}) reveals that there are infinitely many possible labelling choices for the the single-particle motion which preserve the Fourier series of eq.~(\ref{eq:psi_x}). One can show~\cite{mitchell_extracting_2021} that the two formulations: 
\begin{equation}
\left\{
\begin{aligned}
&\vec Q\\
&\vec I\\
&\{\vec n_k,\ \vec m_k,\ \vec \ell_k\}
\end{aligned}
\right.
\ \Leftrightarrow\ 
\left\{
\begin{aligned}
&U \vec Q\\
&(U^T)^{-1}\vec I\\
&(U^T)^{-1} \{\vec n_k,\ \vec m_k,\ \vec \ell_k\}
\end{aligned}
\right.
\label{eq:equivalent_Q}
\end{equation}

\noindent are equivalent when $U$ is taken to be a unimodular (\textit{i.e.} $\det U = \pm 1$) integer matrix. For example, in the case of the high amplitude particle of Fig.~\ref{fig:LHC_continuity}, if one identifies the sideband $Q_x + Q_\zeta$ to be the horizontal fundamental frequency, $Q_x \leftarrow Q_x + Q_\zeta $, the unimodular matrix
$$U = \begin{pmatrix}
    1 & 0 & 1 \\ 0 & 1 & 0 \\ 0 & 0 & 1
\end{pmatrix},\qquad (U^T)^{-1} = \begin{pmatrix}
    1 & 0 & 0 \\ 0 & 1 & 0 \\ -1 & 0 & 1
\end{pmatrix}$$

\noindent shows that indexing the frequencies according to $(n_x,n_y,-n_x+n_\zeta)_k$, \textit{e.g.} the line $(1,0,0)$ becoming $(1,0,-1)$ and $(1,0,1)$ becoming $(1,0,0)$ and so on, yields an equivalent description of the motion through eq.~(\ref{eq:psi_x}). That being said, the three actions are also modified to become $(I_x,I_y,-I_x+I_\zeta)$, which are three new constants of the motion, different (and discontinuous) from the ones that would be expected around the origin. In the same way that the frequencies $\nu_k = \vec n_k \cdot \vec Q$ are preserved under the unimodular transformation $U$, one can show that the dot product:
\begin{equation}
    \vec I' \cdot \vec Q' = \Big[(U^T)^{-1}\vec I\Big]\cdot \Big[U \vec Q\Big] = \vec I \cdot \vec Q
    \label{eq:IQ_preservation}
\end{equation}

\noindent is preserved, independently of the unimodular matrix $U$. Hence, although the mislabelling process can lead to erroneous evaluations of the individual actions, one can always use eq.~(\ref{eq:IQ_preservation}) to identify if the problem stems from this issue, or from the numerical evaluation itself.\\

To lift the equivalence condition of eq.~(\ref{eq:equivalent_Q}), and thereby uniquely label the spectral lines of the single-particle motion, it appears that the only robust method is to invoke the continuity of eq.(\ref{eq:continuity}) by smoothly moving the particle from the origin (where labelling should be unambiguous) to the actual coordinates under study and properly identify $\vec Q$. Although this method is computationally demanding, it is the only one known to the authors to reliably label the spectral lines in complex systems like the LHC, with coupling in all planes. If the particle amplitude is low enough, or if the coupling is less important, one can usually retrieve the fundamental frequencies without worrying about this issue. This is the case for the results shown in Fig.~\ref{fig:LHC_example_ip1_ip5} and Fig.~\ref{fig:LHC_Ix_error}, where the longitudinal coordinate was kept below $\zeta \le 1\sigma_\zeta$ to avoid mislabelling problems.

\section{Hénon map, integral of motion \label{sec:henon_error}}

\begin{figure*}[t!]
    \centering
    ~\hfill
    \subfloat{%
    \centering
    \includegraphics[width=0.49\textwidth]{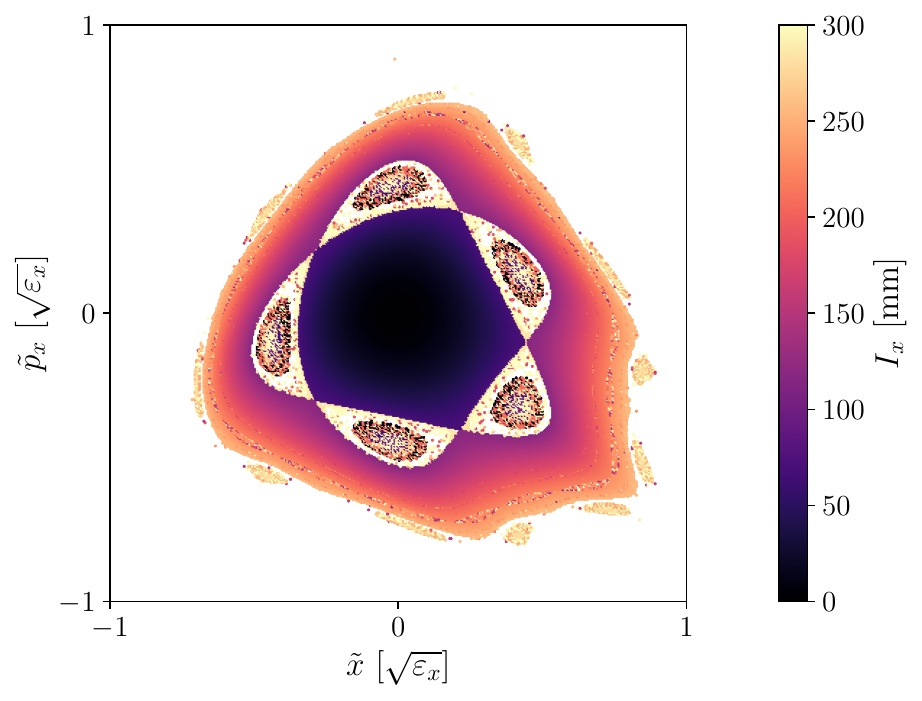}
    }
    ~\hspace{0.015\textwidth}
    \centering
    \subfloat{%
    \includegraphics[width=0.49\textwidth]{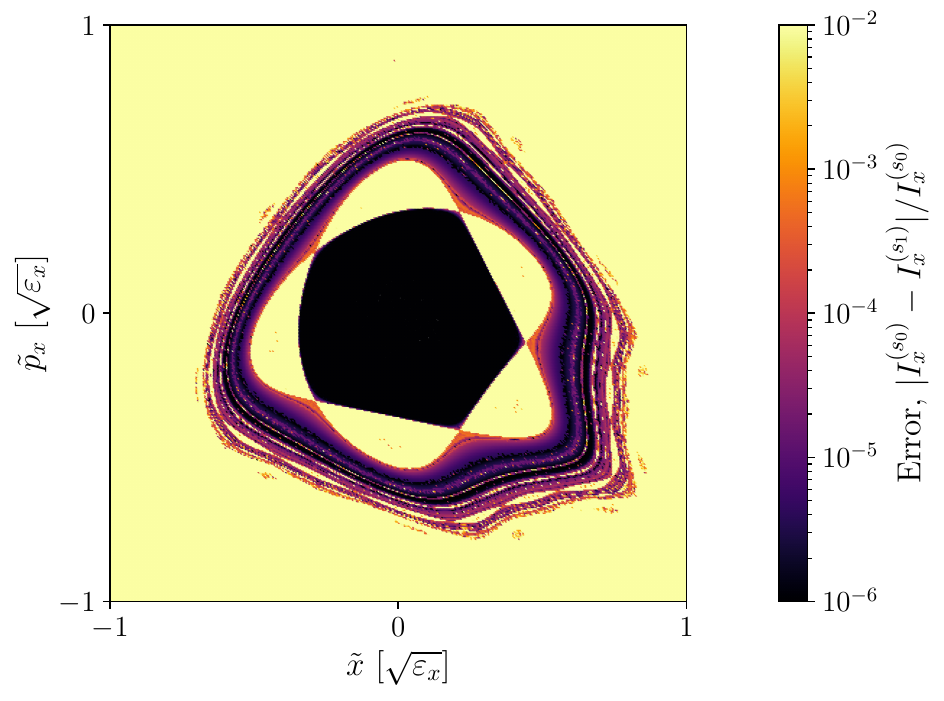}
    }
    \caption{Numerical evaluation of the action, $I_x$, using $N_h=20$ harmonics for different initial conditions in the 2D Hénon map ($\mu_x = 0.2071$). In the right plot, the variation of the action between two observations points, $s_0$ and $s_1$, is shown. By increasing the number of harmonics $N_h$ considered, the picture darkens and all stable tori are evaluated well-below the $10^{-6}$ error value.}
    \label{fig:henon_Ix}
\end{figure*}

Using the procedure described in the paper, one can evaluate the integral of motion, $I_x$, of the 2D Hénon map for various initial conditions in the $(\tilde x, \tilde p_x)$ plane. As shown in Fig.~\ref{fig:henon_Ix} (left), the action continuously increases with the distance from the origin. Within the resonance islands, the action is discontinuous and should instead be evaluated with regards to the appropriate fixed point, located at the center of each island. Additionally, one can assess the invariance of the integral of motion under the Hamiltonian transformation of the map, similar to what was done in Fig.~\ref{fig:LHC_Ix_error} for the LHC. This can be done in time (by evaluating the integral of motion twice, after several turns), or in $s$, by adding additional observations points, as shown in the rightmost plot of Fig~\ref{fig:henon_Ix}. To do so, the sextupolar kick of the Hénon map is split into two equal thin lenses, and the integral of motion is evaluated separately at those two locations, $s_0$ and $s_1$. If more harmonics are considered for the numerical evaluation ($N_h=20$ here), the picture darkens and the error decreases for all initial conditions belonging to KAM tori.




\bibliography{references}

\begin{thebibliography}{33}%
\makeatletter
\providecommand \@ifxundefined [1]{%
 \@ifx{#1\undefined}
}%
\providecommand \@ifnum [1]{%
 \ifnum #1\expandafter \@firstoftwo
 \else \expandafter \@secondoftwo
 \fi
}%
\providecommand \@ifx [1]{%
 \ifx #1\expandafter \@firstoftwo
 \else \expandafter \@secondoftwo
 \fi
}%
\providecommand \natexlab [1]{#1}%
\providecommand \enquote  [1]{``#1''}%
\providecommand \bibnamefont  [1]{#1}%
\providecommand \bibfnamefont [1]{#1}%
\providecommand \citenamefont [1]{#1}%
\providecommand \href@noop [0]{\@secondoftwo}%
\providecommand \href [0]{\begingroup \@sanitize@url \@href}%
\providecommand \@href[1]{\@@startlink{#1}\@@href}%
\providecommand \@@href[1]{\endgroup#1\@@endlink}%
\providecommand \@sanitize@url [0]{\catcode `\\12\catcode `\$12\catcode `\&12\catcode `\#12\catcode `\^12\catcode `\_12\catcode `\%12\relax}%
\providecommand \@@startlink[1]{}%
\providecommand \@@endlink[0]{}%
\providecommand \url  [0]{\begingroup\@sanitize@url \@url }%
\providecommand \@url [1]{\endgroup\@href {#1}{\urlprefix }}%
\providecommand \urlprefix  [0]{URL }%
\providecommand \Eprint [0]{\href }%
\providecommand \doibase [0]{https://doi.org/}%
\providecommand \selectlanguage [0]{\@gobble}%
\providecommand \bibinfo  [0]{\@secondoftwo}%
\providecommand \bibfield  [0]{\@secondoftwo}%
\providecommand \translation [1]{[#1]}%
\providecommand \BibitemOpen [0]{}%
\providecommand \bibitemStop [0]{}%
\providecommand \bibitemNoStop [0]{.\EOS\space}%
\providecommand \EOS [0]{\spacefactor3000\relax}%
\providecommand \BibitemShut  [1]{\csname bibitem#1\endcsname}%
\let\auto@bib@innerbib\@empty
\bibitem [{\citenamefont {Moser}(1978)}]{moser_is_nodate}%
  \BibitemOpen
  \bibfield  {author} {\bibinfo {author} {\bibfnamefont {J.}~\bibnamefont {Moser}},\ }\bibfield  {title} {\bibinfo {title} {Is the solar system stable?},\ }\href@noop {} {\bibfield  {journal} {\bibinfo  {journal} {The Mathematical Intelligencer}\ } (\bibinfo {year} {1978})}\BibitemShut {NoStop}%
\bibitem [{\citenamefont {Poincar{\'e}}(1892)}]{alma9913588812806531}%
  \BibitemOpen
  \bibfield  {author} {\bibinfo {author} {\bibfnamefont {H.}~\bibnamefont {Poincar{\'e}}},\ }\href@noop {} {\emph {\bibinfo {title} {Les m{\'e}thodes nouvelles de la m{\'e}canique c{\'e}leste.}}}\ (\bibinfo {address} {Paris},\ \bibinfo {year} {1892})\BibitemShut {NoStop}%
\bibitem [{\citenamefont {Percival}(1987)}]{percival_chaos_1987}%
  \BibitemOpen
  \bibfield  {author} {\bibinfo {author} {\bibfnamefont {I.~C.}\ \bibnamefont {Percival}},\ }\bibfield  {title} {\bibinfo {title} {Chaos in hamiltonian systems},\ }\href {https://doi.org/10.1098/rspa.1987.0105} {\bibfield  {journal} {\bibinfo  {journal} {Proceedings of the Royal Society of London. A. Mathematical and Physical Sciences}\ }\textbf {\bibinfo {volume} {413}},\ \bibinfo {pages} {131} (\bibinfo {year} {1987})}\BibitemShut {NoStop}%
\bibitem [{\citenamefont {Laskar}(1996)}]{laskar_large_nodate}%
  \BibitemOpen
  \bibfield  {author} {\bibinfo {author} {\bibfnamefont {J.}~\bibnamefont {Laskar}},\ }\bibfield  {title} {\bibinfo {title} {Large scale chaos and marginal stability in the solar system},\ }\href@noop {} {\bibfield  {journal} {\bibinfo  {journal} {Celestial Mechanics and Dynamical Astronomy}\ } (\bibinfo {year} {1996})}\BibitemShut {NoStop}%
\bibitem [{\citenamefont {Wolfram}(2023)}]{wolfram2023}%
  \BibitemOpen
  \bibfield  {author} {\bibinfo {author} {\bibfnamefont {S.}~\bibnamefont {Wolfram}},\ }\href {https://writings.stephenwolfram.com/2023/10/how-to-think-computationally-about-ai-the-universe-and-everything} {\bibinfo {title} {How to think computationally about ai, the universe and everything}},\ \bibinfo {howpublished} {Stephen Wolfram Writings} (\bibinfo {year} {2023})\BibitemShut {NoStop}%
\bibitem [{\citenamefont {Zwirn}\ and\ \citenamefont {Delahaye}(2013)}]{zwirn_unpredictability_2013}%
  \BibitemOpen
  \bibfield  {author} {\bibinfo {author} {\bibfnamefont {H.}~\bibnamefont {Zwirn}}\ and\ \bibinfo {author} {\bibfnamefont {J.-P.}\ \bibnamefont {Delahaye}},\ }\bibfield  {title} {\bibinfo {title} {Unpredictability and {Computational} {Irreducibility}},\ }in\ \href {https://doi.org/10.1007/978-3-642-35482-3_19} {\emph {\bibinfo {booktitle} {Irreducibility and {Computational} {Equivalence}: 10 {Years} {After} {Wolfram}'s {A} {New} {Kind} of {Science}}}}\ (\bibinfo {address} {Berlin, Heidelberg},\ \bibinfo {year} {2013})\ pp.\ \bibinfo {pages} {273--295}\BibitemShut {NoStop}%
\bibitem [{\citenamefont {Courant}\ and\ \citenamefont {Snyder}(1958)}]{courant_theory_1958}%
  \BibitemOpen
  \bibfield  {author} {\bibinfo {author} {\bibfnamefont {E.}~\bibnamefont {Courant}}\ and\ \bibinfo {author} {\bibfnamefont {H.}~\bibnamefont {Snyder}},\ }\bibfield  {title} {\bibinfo {title} {Theory of the alternating-gradient synchrotron},\ }\href {https://doi.org/10.1016/0003-4916(58)90012-5} {\bibfield  {journal} {\bibinfo  {journal} {Annals of Physics}\ }\textbf {\bibinfo {volume} {3}},\ \bibinfo {pages} {1} (\bibinfo {year} {1958})}\BibitemShut {NoStop}%
\bibitem [{\citenamefont {Wolski}(2014)}]{wolski_beam_2014}%
  \BibitemOpen
  \bibfield  {author} {\bibinfo {author} {\bibfnamefont {A.}~\bibnamefont {Wolski}},\ }\href@noop {} {\emph {\bibinfo {title} {Beam dynamics in high energy particle accelerators}}}\ (\bibinfo {year} {2014})\BibitemShut {NoStop}%
\bibitem [{\citenamefont {Chao}(2008)}]{chao_slim_nodate}%
  \BibitemOpen
  \bibfield  {author} {\bibinfo {author} {\bibfnamefont {A.}~\bibnamefont {Chao}},\ }\bibfield  {title} {\bibinfo {title} {{SLIM} - {An} {Early} {Work} {Revisited}},\ }\href@noop {} {\bibfield  {journal} {\bibinfo  {journal} {Particle accelerator. Proceedings, 11th European Conference, EPAC}\ } (\bibinfo {year} {2008})}\BibitemShut {NoStop}%
\bibitem [{\citenamefont {Belanger}\ and\ \citenamefont {Sterbini}()}]{belanger_topo_2024}%
  \BibitemOpen
  \bibfield  {author} {\bibinfo {author} {\bibfnamefont {P.}~\bibnamefont {Belanger}}\ and\ \bibinfo {author} {\bibfnamefont {G.}~\bibnamefont {Sterbini}},\ }\bibfield  {title} {\bibinfo {title} {A topological description of non-linearities in accelerator beam dynamics},\ }\href@noop {} {\bibinfo  {journal} {(in preparation)}\ }\BibitemShut {NoStop}%
\bibitem [{\citenamefont {Gallavotti}(1999)}]{gallavotti_quasi_1999}%
  \BibitemOpen
\bibfield  {journal} {  }\bibfield  {author} {\bibinfo {author} {\bibfnamefont {G.}~\bibnamefont {Gallavotti}},\ }\bibfield  {title} {\bibinfo {title} {Quasi periodic motions from {Hipparchus} to {Kolmogorov}},\ }\href {http://arxiv.org/abs/chao-dyn/9907004} {\bibfield  {journal} {\bibinfo  {journal} {arXiv:1111.4121}\ } (\bibinfo {year} {1999})}\BibitemShut {NoStop}%
\bibitem [{\citenamefont {Bazzani}\ \emph {et~al.}(1994)\citenamefont {Bazzani}, \citenamefont {Todesco}, \citenamefont {Turchetti},\ and\ \citenamefont {Servizi}}]{bazzani_normal_1994-2}%
  \BibitemOpen
  \bibfield  {author} {\bibinfo {author} {\bibfnamefont {A.}~\bibnamefont {Bazzani}}, \bibinfo {author} {\bibfnamefont {E.}~\bibnamefont {Todesco}}, \bibinfo {author} {\bibfnamefont {G.}~\bibnamefont {Turchetti}},\ and\ \bibinfo {author} {\bibfnamefont {G.}~\bibnamefont {Servizi}},\ }\bibfield  {title} {\bibinfo {title} {A normal form approach to the theory of nonlinear betatronic motion}\ }\href {https://doi.org/10.5170/CERN-1994-002} {10.5170/CERN-1994-002} (\bibinfo {year} {1994})\BibitemShut {NoStop}%
\bibitem [{\citenamefont {Franchi}\ \emph {et~al.}(2014)\citenamefont {Franchi}, \citenamefont {Farvacque}, \citenamefont {Ewald}, \citenamefont {Le~Bec},\ and\ \citenamefont {Scheidt}}]{franchi_first_2014}%
  \BibitemOpen
  \bibfield  {author} {\bibinfo {author} {\bibfnamefont {A.}~\bibnamefont {Franchi}}, \bibinfo {author} {\bibfnamefont {L.}~\bibnamefont {Farvacque}}, \bibinfo {author} {\bibfnamefont {F.}~\bibnamefont {Ewald}}, \bibinfo {author} {\bibfnamefont {G.}~\bibnamefont {Le~Bec}},\ and\ \bibinfo {author} {\bibfnamefont {K.}~\bibnamefont {Scheidt}},\ }\bibfield  {title} {\bibinfo {title} {First simultaneous measurement of sextupolar and octupolar resonance driving terms in a circular accelerator from turn-by-turn beam position monitor data},\ }\href {https://doi.org/10.1103/PhysRevSTAB.17.074001} {\bibfield  {journal} {\bibinfo  {journal} {Physical Review Special Topics - Accelerators and Beams}\ }\textbf {\bibinfo {volume} {17}},\ \bibinfo {pages} {074001} (\bibinfo {year} {2014})}\BibitemShut {NoStop}%
\bibitem [{\citenamefont {Forest}(2006)}]{forest_geometric_2006-1}%
  \BibitemOpen
  \bibfield  {author} {\bibinfo {author} {\bibfnamefont {E.}~\bibnamefont {Forest}},\ }\bibfield  {title} {\bibinfo {title} {Geometric integration for particle accelerators},\ }\href {https://doi.org/10.1088/0305-4470/39/19/S03} {\bibfield  {journal} {\bibinfo  {journal} {Journal of Physics A: Mathematical and General}\ }\textbf {\bibinfo {volume} {39}},\ \bibinfo {pages} {5321} (\bibinfo {year} {2006})}\BibitemShut {NoStop}%
\bibitem [{\citenamefont {Dragt}(2013)}]{dragt_overview_2013-1}%
  \BibitemOpen
  \bibfield  {author} {\bibinfo {author} {\bibfnamefont {A.~J.}\ \bibnamefont {Dragt}},\ }\bibfield  {title} {\bibinfo {title} {An {Overview} of {Lie} {Methods} for {Accelerator} {Physics}},\ }\href@noop {} {\bibfield  {journal} {\bibinfo  {journal} {Proceedings of PAC}\ } (\bibinfo {year} {2013})}\BibitemShut {NoStop}%
\bibitem [{\citenamefont {Bazzani}\ \emph {et~al.}(1997)\citenamefont {Bazzani}, \citenamefont {Bongini},\ and\ \citenamefont {Turchetti}}]{bazzani_analysis_1997}%
  \BibitemOpen
  \bibfield  {author} {\bibinfo {author} {\bibfnamefont {A.}~\bibnamefont {Bazzani}}, \bibinfo {author} {\bibfnamefont {L.}~\bibnamefont {Bongini}},\ and\ \bibinfo {author} {\bibfnamefont {G.}~\bibnamefont {Turchetti}},\ }\bibfield  {title} {\bibinfo {title} {Analysis of resonances by action map comparing tracking and normal forms},\ }in\ \href {https://doi.org/10.1063/1.52927} {\emph {\bibinfo {booktitle} {{AIP} {Conference} {Proceedings}}}}\ (\bibinfo {year} {1997})\BibitemShut {NoStop}%
\bibitem [{\citenamefont {Meiss}(1992)}]{meiss_symplectic_1992}%
  \BibitemOpen
  \bibfield  {author} {\bibinfo {author} {\bibfnamefont {J.~D.}\ \bibnamefont {Meiss}},\ }\bibfield  {title} {\bibinfo {title} {Symplectic maps, variational principles, and transport},\ }\href {https://doi.org/10.1103/RevModPhys.64.795} {\bibfield  {journal} {\bibinfo  {journal} {Reviews of Modern Physics}\ }\textbf {\bibinfo {volume} {64}},\ \bibinfo {pages} {795} (\bibinfo {year} {1992})}\BibitemShut {NoStop}%
\bibitem [{\citenamefont {Hénon}(1969)}]{Henon_numerical_1969}%
  \BibitemOpen
  \bibfield  {author} {\bibinfo {author} {\bibfnamefont {M.}~\bibnamefont {Hénon}},\ }\bibfield  {title} {\bibinfo {title} {Numerical study of quadratic area-preserving mappings},\ }\href {https://doi.org/10.1090/qam/253513} {\bibfield  {journal} {\bibinfo  {journal} {Quarterly of Applied Mathematics}\ }\textbf {\bibinfo {volume} {27}},\ \bibinfo {pages} {291} (\bibinfo {year} {1969})}\BibitemShut {NoStop}%
\bibitem [{\citenamefont {Henon}\ and\ \citenamefont {Heiles}(1964)}]{Henon_applicability_1964}%
  \BibitemOpen
  \bibfield  {author} {\bibinfo {author} {\bibfnamefont {M.}~\bibnamefont {Henon}}\ and\ \bibinfo {author} {\bibfnamefont {C.}~\bibnamefont {Heiles}},\ }\bibfield  {title} {\bibinfo {title} {The applicability of the third integral of motion: {Some} numerical experiments},\ }\href {https://doi.org/10.1086/109234} {\bibfield  {journal} {\bibinfo  {journal} {The Astronomical Journal}\ }\textbf {\bibinfo {volume} {69}},\ \bibinfo {pages} {73} (\bibinfo {year} {1964})}\BibitemShut {NoStop}%
\bibitem [{\citenamefont {Iadarola}\ \emph {et~al.}(2023)\citenamefont {Iadarola} \emph {et~al.}}]{cern_xsuite_nodate}%
  \BibitemOpen
  \bibfield  {author} {\bibinfo {author} {\bibfnamefont {G.}~\bibnamefont {Iadarola}} \emph {et~al.},\ }\href {https://github.com/xsuite/xsuite/blob/main/docs/physics_manual/physics_man.pdf} {\emph {\bibinfo {title} {{Xsuite Physics Manual}}}}\ (\bibinfo {year} {2023})\ \bibinfo {note} {https://xsuite.web.cern.ch}\BibitemShut {NoStop}%
\bibitem [{\citenamefont {Bartolini}\ and\ \citenamefont {Schmidt}(1998)}]{bartolini_normal_1997}%
  \BibitemOpen
  \bibfield  {author} {\bibinfo {author} {\bibfnamefont {R.}~\bibnamefont {Bartolini}}\ and\ \bibinfo {author} {\bibfnamefont {F.}~\bibnamefont {Schmidt}},\ }\bibfield  {title} {\bibinfo {title} {{Normal form via tracking or beam data}},\ }\href@noop {} {\bibfield  {journal} {\bibinfo  {journal} {Part. Accel.}\ }\textbf {\bibinfo {volume} {59}},\ \bibinfo {pages} {93} (\bibinfo {year} {1998})}\BibitemShut {NoStop}%
\bibitem [{\citenamefont {Laskar}(1999)}]{simo_introduction_1999}%
  \BibitemOpen
  \bibfield  {author} {\bibinfo {author} {\bibfnamefont {J.}~\bibnamefont {Laskar}},\ }\bibfield  {title} {\bibinfo {title} {Introduction to {Frequency} {Map} {Analysis}},\ }in\ \href {https://doi.org/10.1007/978-94-011-4673-9_13} {\emph {\bibinfo {booktitle} {Hamiltonian {Systems} with {Three} or {More} {Degrees} of {Freedom}}}},\ \bibinfo {editor} {edited by\ \bibinfo {editor} {\bibfnamefont {C.}~\bibnamefont {Simó}}}\ (\bibinfo  {publisher} {Springer},\ \bibinfo {year} {1999})\ pp.\ \bibinfo {pages} {134--150}\BibitemShut {NoStop}%
\bibitem [{\citenamefont {Belanger}\ \emph {et~al.}(2024{\natexlab{a}})\citenamefont {Belanger}, \citenamefont {Paraschou}, \citenamefont {Sterbini}, \citenamefont {Iadarola} \emph {et~al.}}]{nafflib_pypi}%
  \BibitemOpen
  \bibfield  {author} {\bibinfo {author} {\bibfnamefont {P.}~\bibnamefont {Belanger}}, \bibinfo {author} {\bibfnamefont {K.}~\bibnamefont {Paraschou}}, \bibinfo {author} {\bibfnamefont {G.}~\bibnamefont {Sterbini}}, \bibinfo {author} {\bibfnamefont {G.}~\bibnamefont {Iadarola}}, \emph {et~al.},\ }\href {https://pypi.org/project/nafflib/} {\bibinfo {title} {nafflib: {A Python Implementation of the Numerical Analysis of the Fundamental Frequencies}}} (\bibinfo {year} {2024}{\natexlab{a}}),\ \bibinfo {note} {https://pypi.org/project/nafflib/}\BibitemShut {NoStop}%
\bibitem [{\citenamefont {Laskar}(1993)}]{laskar_frequency_1993}%
  \BibitemOpen
  \bibfield  {author} {\bibinfo {author} {\bibfnamefont {J.}~\bibnamefont {Laskar}},\ }\bibfield  {title} {\bibinfo {title} {Frequency analysis for multi-dimensional systems. {Global} dynamics and diffusion},\ }\href {https://doi.org/10.1016/0167-2789(93)90210-R} {\bibfield  {journal} {\bibinfo  {journal} {Physica D: Nonlinear Phenomena}\ }\textbf {\bibinfo {volume} {67}},\ \bibinfo {pages} {257} (\bibinfo {year} {1993})}\BibitemShut {NoStop}%
\bibitem [{\citenamefont {Laskar}\ \emph {et~al.}(1992)\citenamefont {Laskar}, \citenamefont {Froeschlé},\ and\ \citenamefont {Celletti}}]{laskar_measure_1992}%
  \BibitemOpen
  \bibfield  {author} {\bibinfo {author} {\bibfnamefont {J.}~\bibnamefont {Laskar}}, \bibinfo {author} {\bibfnamefont {C.}~\bibnamefont {Froeschlé}},\ and\ \bibinfo {author} {\bibfnamefont {A.}~\bibnamefont {Celletti}},\ }\bibfield  {title} {\bibinfo {title} {The {Measure} of {Chaos} by the {Numerical} {Analysis} of the {Fundamental} {Frequencies}. {Application} to the {Standard} {Mapping}},\ }\href {https://doi.org/10.1016/0167-2789(92)90028-L} {\bibfield  {journal} {\bibinfo  {journal} {Physica D: Nonlinear Phenomena}\ }\textbf {\bibinfo {volume} {56}},\ \bibinfo {pages} {253} (\bibinfo {year} {1992})}\BibitemShut {NoStop}%
\bibitem [{\citenamefont {Laskar}(2003)}]{laskar_frequency_2003}%
  \BibitemOpen
  \bibfield  {author} {\bibinfo {author} {\bibfnamefont {J.}~\bibnamefont {Laskar}},\ }\bibfield  {title} {\bibinfo {title} {Frequency map analysis and particle accelerators},\ }in\ \href {https://doi.org/10.1109/PAC.2003.1288929} {\emph {\bibinfo {booktitle} {Proceedings of the 2003 {Bipolar}/{BiCMOS} {Circuits} and {Technology} {Meeting}}}}\ (\bibinfo {year} {2003})\ pp.\ \bibinfo {pages} {378--382}\BibitemShut {NoStop}%
\bibitem [{\citenamefont {Papaphilippou}(2014)}]{papaphilippou_detecting_2014}%
  \BibitemOpen
  \bibfield  {author} {\bibinfo {author} {\bibfnamefont {Y.}~\bibnamefont {Papaphilippou}},\ }\bibfield  {title} {\bibinfo {title} {Detecting chaos in particle accelerators through the frequency map analysis method},\ }\href {https://doi.org/10.1063/1.4884495} {\bibfield  {journal} {\bibinfo  {journal} {Chaos: An Interdisciplinary Journal of Nonlinear Science}\ }\textbf {\bibinfo {volume} {24}},\ \bibinfo {pages} {024412} (\bibinfo {year} {2014})}\BibitemShut {NoStop}%
\bibitem [{\citenamefont {Masoliver}\ and\ \citenamefont {Ros}(2011)}]{masoliver_integrability_2011}%
  \BibitemOpen
  \bibfield  {author} {\bibinfo {author} {\bibfnamefont {J.}~\bibnamefont {Masoliver}}\ and\ \bibinfo {author} {\bibfnamefont {A.}~\bibnamefont {Ros}},\ }\bibfield  {title} {\bibinfo {title} {Integrability and chaos: the classical uncertainty},\ }\href {https://doi.org/10.1088/0143-0807/32/2/016} {\bibfield  {journal} {\bibinfo  {journal} {European Journal of Physics}\ }\textbf {\bibinfo {volume} {32}},\ \bibinfo {pages} {431} (\bibinfo {year} {2011})}\BibitemShut {NoStop}%
\bibitem [{\citenamefont {Herr}\ and\ \citenamefont {Pieloni}(2014)}]{herr_beam-beam_2014-1}%
  \BibitemOpen
  \bibfield  {author} {\bibinfo {author} {\bibfnamefont {W.}~\bibnamefont {Herr}}\ and\ \bibinfo {author} {\bibfnamefont {T.}~\bibnamefont {Pieloni}},\ }\bibfield  {title} {\bibinfo {title} {Beam-{Beam} {Effects}},\ }in\ \href {https://doi.org/10.5170/CERN-2014-009.431} {\emph {\bibinfo {booktitle} {{CAS} - {CERN} {Accelerator} {School}: {Advanced} {Accelerator} {Physics}}}}\ (\bibinfo {year} {2014})\ pp.\ \bibinfo {pages} {431--459}\BibitemShut {NoStop}%
\bibitem [{\citenamefont {Belanger}\ \emph {et~al.}(2024{\natexlab{b}})\citenamefont {Belanger}, \citenamefont {Baartman}, \citenamefont {Kaltchev}, \citenamefont {Iadarola},\ and\ \citenamefont {Sterbini}}]{belanger_bunch-by-bunch_2024-2}%
  \BibitemOpen
  \bibfield  {author} {\bibinfo {author} {\bibfnamefont {P.}~\bibnamefont {Belanger}}, \bibinfo {author} {\bibfnamefont {R.}~\bibnamefont {Baartman}}, \bibinfo {author} {\bibfnamefont {D.}~\bibnamefont {Kaltchev}}, \bibinfo {author} {\bibfnamefont {G.}~\bibnamefont {Iadarola}},\ and\ \bibinfo {author} {\bibfnamefont {G.}~\bibnamefont {Sterbini}},\ }\bibfield  {title} {\bibinfo {title} {Bunch-by-bunch simulations of beam-beam driven particle losses in the {LHC}},\ }\href@noop {} {\bibfield  {journal} {\bibinfo  {journal} {IPAC24}\ } (\bibinfo {year} {2024}{\natexlab{b}})}\BibitemShut {NoStop}%
\bibitem [{\citenamefont {Hirata}\ \emph {et~al.}(2003)\citenamefont {Hirata}, \citenamefont {Moshammer},\ and\ \citenamefont {Ruggiero}}]{hirata_symplectic_2003}%
  \BibitemOpen
  \bibfield  {author} {\bibinfo {author} {\bibfnamefont {K.}~\bibnamefont {Hirata}}, \bibinfo {author} {\bibfnamefont {H.}~\bibnamefont {Moshammer}},\ and\ \bibinfo {author} {\bibfnamefont {F.}~\bibnamefont {Ruggiero}},\ }\href {https://doi.org/10.2172/813308} {\emph {\bibinfo {title} {A {Symplectic} {Beam}-{Beam} {Interaction} with {Energy} {Change}}}},\ \bibinfo {type} {Tech. Rep.}\ \bibinfo {number} {SLAC-PUB-10055, 813308}\ (\bibinfo {year} {2003})\BibitemShut {NoStop}%
\bibitem [{\citenamefont {Iadarola}\ \emph {et~al.}(2024)\citenamefont {Iadarola} \emph {et~al.}}]{Iadarola:2023fuk}%
  \BibitemOpen
  \bibfield  {author} {\bibinfo {author} {\bibfnamefont {G.}~\bibnamefont {Iadarola}} \emph {et~al.},\ }\bibfield  {title} {\bibinfo {title} {{Xsuite: An Integrated Beam Physics Simulation Framework}},\ }\bibfield  {journal} {\bibinfo  {journal} {JACoW}\ }\href {https://doi.org/10.18429/JACoW-HB2023-TUA2I1} {10.18429/JACoW-HB2023-TUA2I1} (\bibinfo {year} {2024})\BibitemShut {NoStop}%
\bibitem [{\citenamefont {Mitchell}\ \emph {et~al.}(2021)\citenamefont {Mitchell}, \citenamefont {Ryne}, \citenamefont {Hwang}, \citenamefont {Nagaitsev},\ and\ \citenamefont {Zolkin}}]{mitchell_extracting_2021}%
  \BibitemOpen
  \bibfield  {author} {\bibinfo {author} {\bibfnamefont {C.~E.}\ \bibnamefont {Mitchell}}, \bibinfo {author} {\bibfnamefont {R.~D.}\ \bibnamefont {Ryne}}, \bibinfo {author} {\bibfnamefont {K.}~\bibnamefont {Hwang}}, \bibinfo {author} {\bibfnamefont {S.}~\bibnamefont {Nagaitsev}},\ and\ \bibinfo {author} {\bibfnamefont {T.}~\bibnamefont {Zolkin}},\ }\bibfield  {title} {\bibinfo {title} {Extracting {Dynamical} {Frequencies} from {Invariants} of {Motion} in {Finite}-{Dimensional} {Nonlinear} {Integrable} {Systems}},\ }\bibfield  {journal} {\bibinfo  {journal} {Physical Review E}\ }\textbf {\bibinfo {volume} {103}},\ \href {https://doi.org/10.1103/PhysRevE.103.062216} {10.1103/PhysRevE.103.062216} (\bibinfo {year} {2021})\BibitemShut {NoStop}%
\end{thebibliography}%

\end{document}